\documentclass[11pt]{article}
\usepackage[utf8]{inputenc}
\usepackage{tocloft} 
\usepackage{amsmath}
\usepackage{amssymb}
\usepackage{graphicx}
\usepackage{multirow}

\usepackage[table,dvipsnames]{xcolor}

\usepackage{threeparttable}
\usepackage{adjustbox}

\usepackage{booktabs} 
\usepackage{tabularx}
\usepackage{floatrow}
\usepackage{scrextend}
\usepackage{csquotes}

\usepackage{caption}
\usepackage{subcaption}

\usepackage[margin=2.8cm]{geometry}
\usepackage[textsize=footnotesize]{todonotes}
\usepackage{authblk}

\usepackage{xcolor}
\usepackage[colorlinks = true,
            linkcolor = blue,
            urlcolor  = blue,
            citecolor = blue,
            anchorcolor = blue]{hyperref}
\usepackage{pdflscape}
\usepackage{appendix} 
\usepackage{placeins} 
\usepackage{fancyhdr}
\usepackage[backend=biber,
    style=numeric-comp,
    sorting=none,
    doi=false,
    isbn=false,
    url=false,
    eprint=false]{biblatex}
\appto{\bibsetup}{\sloppy} 

\title{Trading off performance and human oversight in algorithmic policy: evidence from Danish college admissions\footnote{We are grateful for comments that we have received in presentations at MD4SG/EAAMO, Copenhagen Education Network, Social Complexity Lab, DTU and CPH Tech Policy Committee as well as from Rene Kizilcec,  Sune Lehmann, Mikkel H. Gandil and Tereza Blazkova. This project was funded by the Nation-Scale Social Networks grant from the Villum Foundation, a seed grant from Economic Policy Research Network and Independent Research Foundation Denmark grant 3099-00139B. All authors declare no competing interests. Views and conclusions expressed in the article are those of the authors and do not necessarily represent those of the Rectorate of University of Copenhagen. }}
\author[1,2,*]{Magnus Lindgaard Nielsen}
\author[1,2]{Jonas Skjold Raaschou-Pedersen}
\author[1]{Emil Chrisander}
\author[1]{David Dreyer Lassen}
\author[3]{Julien Grenet}
\author[1,4]{Anna Rogers}
\author[1,*]{Andreas Bjerre-Nielsen}
\affil[1]{University of Copenhagen} 
\affil[2]{Data Science Lab, Statistics Denmark}
\affil[3]{CNRS and Paris School of Economics}
\affil[4]{IT University of Copenhagen}
\affil[*]{Corresponding authors, email addresses are respectively:\\
\href{magnus.nielsen@sodas.ku.dk}{magnus.nielsen@sodas.ku.dk}; \href{mailto:abn@sodas.ku.dk}{abn@sodas.ku.dk} }
\date{\today}

\makeindex

\begin{document}

\maketitle

\captionsetup[figure]{list=no}
\captionsetup[table]{list=no}
\begin{abstract}
\noindent Student dropout is a significant concern for educational institutions due to its social and economic impact, driving the need for risk prediction systems to identify at-risk students before enrollment. We explore the accuracy of such systems in the context of higher education by predicting degree completion before admission, with potential applications for prioritizing admissions decisions. Using a large-scale dataset from Danish higher education admissions, we demonstrate that ML models offer more precise and fair predictions compared to current practices that rely on either high school grade point averages or human judgment. We find that advanced models outperform simpler models, but that these gains are small relative to the gains from replacing current policies with any machine learning model. This implies a trade-off in the performance and transparency of the prediction policy from using more complex models. We estimate that the use of even simple ML models to guide admissions decisions—particularly in response to a newly implemented nationwide policy reducing admissions by 10\%—could yield significant economic benefits. However, this improvement would come at the cost of reduced human oversight and lower transparency. We uniquely combine estimates of potential benefits with discussions of regulatory challenges related to high-risk AI in educational admissions, underscoring the opportunities and complexities of applying machine learning to policy decisions.        
\end{abstract}

\pagebreak

\section{Introduction}
In the context of educational policy, data-driven insights are instrumental in addressing significant challenges, particularly in managing student dropout in traditional on-campus settings (see e.g., \cite{lakkaraju_machine_2015, aulck_predicting_2017, glandorf_temporal_2024}) and online environments (see e.g., \cite{chen_systematic_2024}).
Student dropout carries significant costs at the individual, institutional, and societal levels, depending on educational and institutional arrangements \cite{commission_dropout_2015, liu_reimagining_2023}. 
A key factor contributing to dropout is the potential mismatch between students and their chosen majors, which is exacerbated when students apply to specific fields without the possibility of changing later on \cite{behr_dropping_2020, stinebrickner_major_2014}.  In Danish higher education, dropout rates of 25-30\% have been a persistent concern for both colleges and universities, which are financially impacted due to funding mechanisms that reward timely graduation per student of 21-46 thousand USD, see Section \ref{sec:revenue-calculation} for details. This challenge is mirrored in admission systems globally, where admission officers prioritize applicants using a mix of assessment metrics such as GPA, test scores, and character assessments. These metrics aim to reflect merit and predict future life outcomes, albeit with varying degrees of success \cite{oliveri_assessments_2020, borghans_what_2016}.

One proposed solution is the use of early warning systems, as recent studies have found that the deployment of predictive methods to match students with courses and classrooms at their proficiency level can improve educational outcomes \cite{bergman_seven-college_2021, perdomo_difficult_2023}. 
Data-driven predictions could also improve admissions in the first place, complementing and reducing the need for costly and sensitive post-admission data collection that raises privacy concerns \cite{bjerre-nielsen_task-specific_2021}.

More broadly, advances in machine learning (ML), particularly through large language models \cite{vaswani_attention_2017, devlin_bert_2018, openai_introducing_2023}, offer novel data processing opportunities for public policy. 
In the fields of market and mechanism design, which focus on developing allocation systems like admissions, the potential of data-driven prioritization remains largely underexplored. However, the ``black box'' nature of such models also introduces new risks, emphasizing the need for increased oversight and regulation due to growing public concerns \cite{veale_demystifying_2021, mokander_us_2022}.

Predicting student dropout is challenging but essential, and it has been explored using various data sources and models \cite{alyahyan_predicting_2020}. In the context of admissions, however, research has struggled to identify predictive signals beyond prior grades or legally and ethically prohibited attributes, such as demographic attributes \cite{bjerre-nielsen_task-specific_2021, belfield_predicting_2012, al-alawi_systematic_2020, li_predictive_2023}. Transformer models, known for their ability to process high-granularity sequential information \cite{vaswani_attention_2017}, present a promising solution for more accurate predictions in individual life outcomes, including education, health, and labor \cite{choi_towards_2020, li_behrt_2020, vafa_career_2022, savcisens_using_2024}.  However, their potential to improve decision-making or policies remains underexplored and comes with challenges to transparency and accountability.
Algorithmic models have been shown to reduce decision errors compared to humans, leading to positive net government revenue, thus being self-financing \cite{ludwig_unreasonable_2024}.  
An essential consideration is the equity of policies and how they may discriminate against different sociodemographic groups. Predictive algorithms can be adjusted to reduce disparities in decision-making across demographic groups, potentially outperforming human decision-makers in terms of fairness \cite{hardt_equality_2016, kleinberg_human_2018}. Despite this, current and new algorithmic policies have yet to be widely compared in terms of fairness, independent of decision thresholds.
Moreover, in the context of admissions and allocation systems, existing studies have primarily focused on replicating human assessments of applicants or predicting academic performance in a limited number of programs \cite{waters_grade_2014, timer_use_2011, al-alawi_systematic_2020, li_predictive_2023, lee_evaluating_2023}. Finally, these existing studies have not addressed the trade-offs related to compliance with new regulations, such as the EU AI Act or US Algorithmic Accountability Act, which emphasize transparency and human oversight. 

As a case study, we leverage a unique, nation-wide historical dataset on higher education admissions in Denmark available to researchers through Statistics Denmark (see \ref{sec:data_sources_ethical_considerations} for details), to evaluate whether prediction algorithms can improve admission decisions on a national scale. Our data presents an ideal case due to the high quality, large sample size and diverse types of information captured. Our evaluation focuses on optimally ranking students based on available data by implementing a straightforward prediction policy, where model-generated risk scores create hypothetical rankings of students \cite{kleinberg_prediction_2015, lakkaraju_machine_2015}. The predictive models utilize only pre-admission grades as inputs, generating new, potentially non-linear transformations of grade data that capture more nuanced patterns beyond a simple point average. Due to the large amount of high quality data, we are able to assess both whether simple machine learning models, specifically logistic regression and gradient boosted trees \cite{chen_xgboost_2016}, and more advanced deep learning models, specifically LSTM \cite{hochreiter_long_1997} and the transformer \cite{vaswani_attention_2017}, are able to capture these patterns. 

\begin{figure}
    \centering
    \begin{adjustbox}{center}
    \includegraphics[width=1\textwidth]{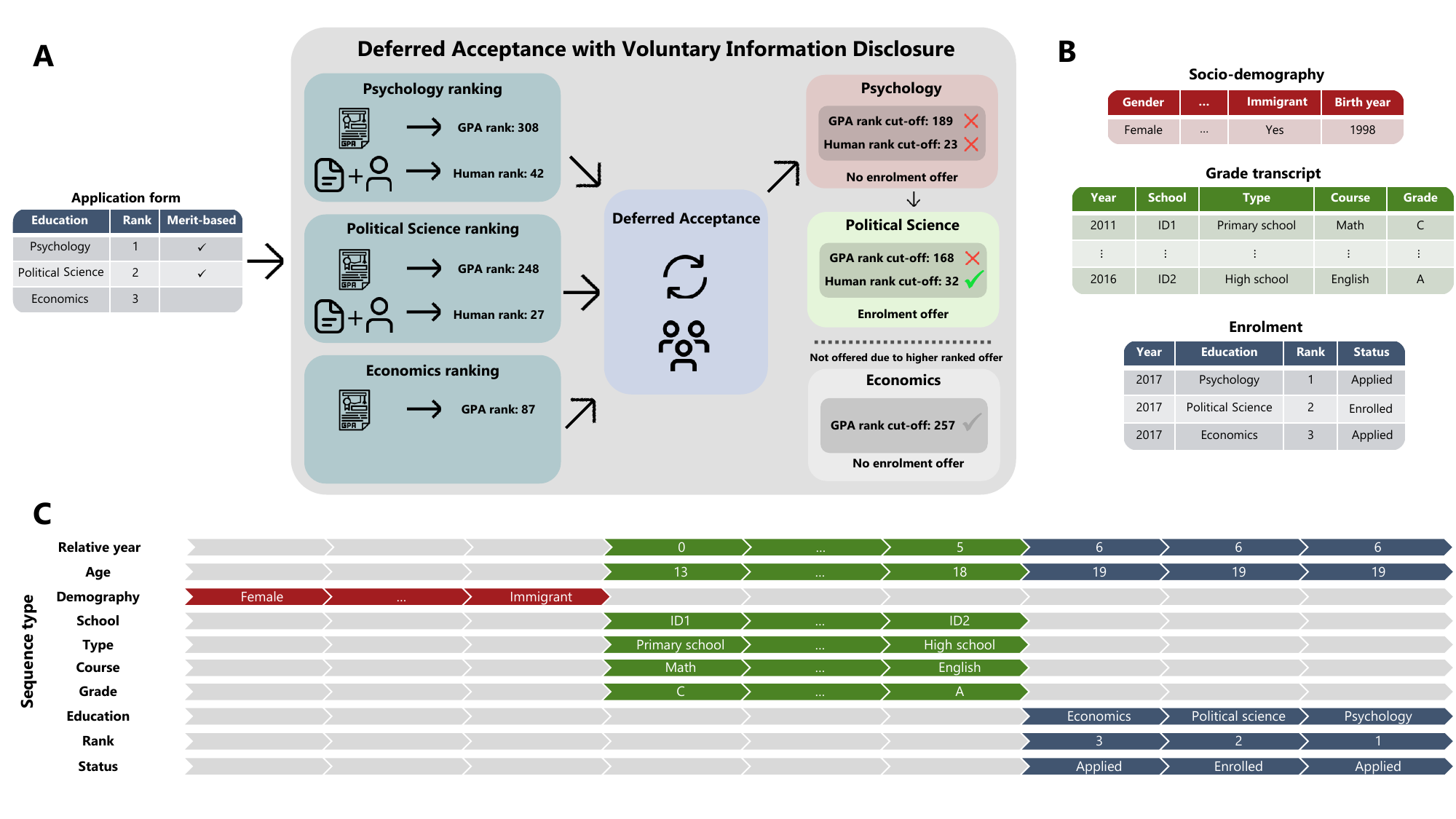}
    \end{adjustbox}
 \caption[Admission to higher education and sequence representation]{\textbf{Admission to higher education and sequence representation.} Panel~A illustrates the general process for admission to higher education in Denmark, depicting two systems for ranking applicants: a mandatory GPA ranking, where all students are ranked by their high school GPA, and a merit-based human ranking, where students can opt-in voluntarily. Both rankings are used as inputs in a variant of the deferred acceptance mechanism \cite{bjerre-nielsen_voluntary_2022}. Student receives a single enrollment offer from the highest-ranked institution where they qualify. Panel~B displays the data observed after student enrollment, showing how information is stored in tabular form. Panel~C shows the creation of 10 sequences based on the tabular representations from panel B, where the different colors indicate the source of the information in the tabular representation. These sequences begin with time-invariant events (e.g., socio-demographic information) followed by chronologically ordered events (e.g., grades). Each sequence encapsulates information regarding a specific aspect of events, represented by a column in the tabular representation, with the \(i\)\textsuperscript{th} position in each sequence describing the \(i\)\textsuperscript{th} event, regardless of its relevance.}
    \label{fig:combined-illustration}
\end{figure}

In centralized admission systems like Denmark's, these rankings directly determine which students are admitted to specific programs (Fig.~\ref{fig:combined-illustration}A). Because centralized admission procedures, such as Denmark's, exclusively use rankings to determine admissions, we can directly compare actual admitted-student rankings under current policies with counterfactual rankings produced by algorithmic predictions—avoiding the selective labels problem faced by prior studies \cite{kleinberg_algorithmic_2018}.

Yet, comparing outcomes among admitted students requires an assumption of no impact on switching between study programs, i.e., general equilibrium effects, a point which we expand upon on in the discussion. Applicants are ranked based on high school GPA, but applicants can also opt-in to admission through a secondary quota where rankings are generated by an assessment performed by humans \cite{gandil_college_2022, bjerre-nielsen_voluntary_2022}. This dual approach (Fig.~\ref{fig:combined-illustration}A) accommodates a diverse applicant pool and can improve graduation rates among those admitted through the secondary quota \cite{gandil_college_2022, bjerre-nielsen_voluntary_2022}, as can also be seen in Table~\ref{tab:summary_stats_quota_1_and_2}, with students admitted through the secondary quota being both older, having lower grade point averages and higher graduation rates, see Section~\ref{sec:institutional_setting} for a further description of the institutional setting. Importantly, this means that our model versus human screening comparison is not for the general student body, but for those who selectively opt in to secondary screening. By restricting our models to input features that are already accessible through the central registers, and focusing on predictions for future cohorts, we provide a credible evaluation of feasible admission policies based on data that is readily available before enrollment \cite{bjerre-nielsen_task-specific_2021, yu_should_2021, liu_reimagining_2023}.

\FloatBarrier
\section{Results}\label{sec:results}

We develop predictive models with study completion as the target variable, using various subsets of input data related to Danish students. These data subsets include raw grade transcripts from middle and high school, sociodemographic information about the students and their parents, application details, human assessments (when applicable), and records of study enrollment and completion.
Our two primary models are based on distinct input sets: (1)~the \textit{academic} set, which includes only pre-admission grades and enrollment details, focusing solely on objective academic criteria permitted by law while excluding protected attributes \cite{uddannelses-_og_forskningsministeriet_bekendtgorelse_2022}; and (2)~the \textit{everything} set, which incorporates all available information, including sociodemographic data, application details, and human assessments (see Section~\ref{sec:input_variants} for details).

We evaluate the performance of two deep learning architectures, the transformer \cite{vaswani_attention_2017} and Long Short-Term Memory (LSTM) \cite{hochreiter_long_1997}, comparing them against each other and against tabular models which do not use sequential data, such as logistic regression and gradient-boosted trees (specifically XGBoost) \cite{chen_xgboost_2016}. Panels~B and~C of Figure~\ref{fig:combined-illustration} illustrate how student information is transformed into sequences for input into these advanced models, with model and training details given in \ref{sec:models}.

\subsection{Model performance}

\begin{figure}[H]
    \centering
        \includegraphics[width=\linewidth]{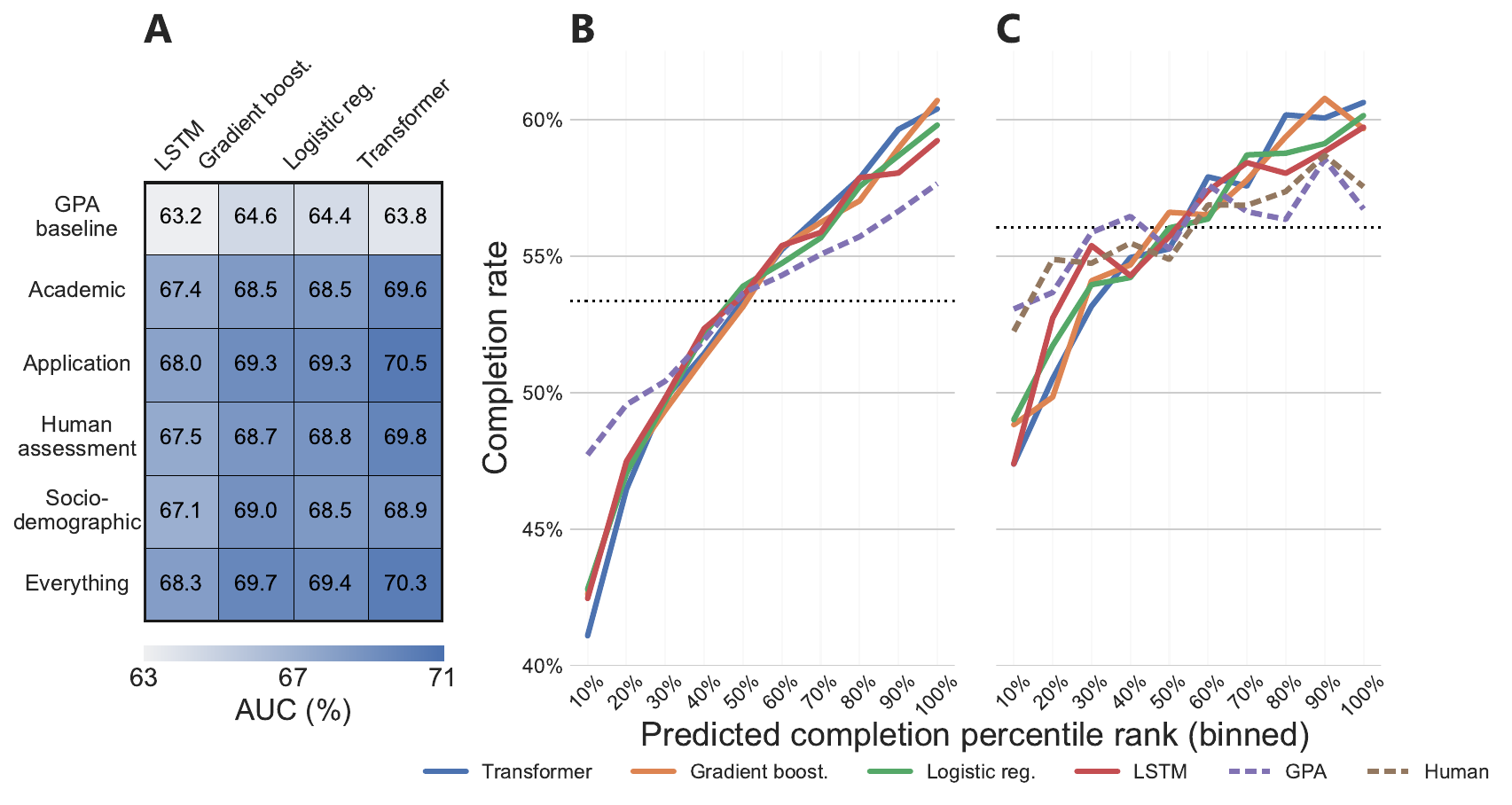}
    \caption[Predictive performance of models and admission criteria]{\textbf{Predictive performance of models and admission criteria.} Panel~A shows the model performance measured by AUC scores for different input types (in rows) and models (in columns). Panels~B and~C display the actual student completion rates as a function of predicted completion rates. The rates are binned by percent deciles, and performance is shown by model rank (solid lines) or observed admission rank (dashed lines). The performance scores are split by admission method: Panel~B for GPA-based admissions and Panel~C for human assessment. Students are perfectly ranked if completion rates start at 0\% and abruptly jump to 100\%. Horizontal dotted lines indicate the mean completion rate within each sample.}
    \label{fig:contract_main}
\end{figure}

We begin by reviewing the performance of all models using the AUC (Area Under the ROC Curve) score \cite{huang_using_2005}, as we are primarily interested in the rankings of students, due to these being used for seat allocation and being comparable to the currently implemented ranking methods (see Panel~A of Figure~\ref{fig:contract_main}). For the \textit{academic} input set, the transformer model achieves the highest AUC score of 69.6\% on our out-of-sample admission data, while the older LSTM architecture yields 67.4\%. In comparison, ranking students based solely on their grade point averages (GPA) and study program indicators---included to encode idiosyncratic variation in graduation rates across study programs---results in AUC scores of up to 64.6\%. Using aggregate grade information yields AUC scores of 68.5\% for both logistic regression and gradient-boosted trees. Thus, any ML algorithm using the \textit{academic} input set to rank students outperforms the current methods of ranking by GPA or human assessment, with the simple logistic regression model showing a 4.1 percentage-point (pp) improvement compared to a baseline using only GPA and place of enrollment, whereas the transformer only shows a 1.1 pp improvement in AUC compared to the logistic regression (see standard errors in Table~\ref{tab:auc}). As such, the largest performance gain stems from using any machine learning model, rather than using advanced sequential deep learning models over simpler tabular models.

We observe some differences in performance based on the input sets used for the models, but the variation in performance across input sets is modest, with models at most reaching AUC scores of 70.5\%. We note that the gains from using \emph{everything} as input in the less advanced models yields a performance gain of up to 1.2 pp points, which is approximately equal the gain from adopting the more advanced architecture, see Table~\ref{tab:auc} where the s.e. for both estimates are 0.3 pp.
We also note that one of the core information sources in admissions is the \emph{application} feature set, i.e., the program's applicants applied to and the order of their preferences. This feature set provides valuable insights into applicants' ambitions and perceived skills, leading to the best-performing model when integrated. This highlights a fundamental tension between prediction quality and manipulability, a topic which is revisited in the Discussion, Section \ref{sec:discussion}. Another potential concern with deep learning models is the instability of their performance across different initializations \cite{mosbach_stability_2021}. However, we find that our performance measures remain relatively stable, with standard errors across initializations at or below 0.3 pp (see Table~\ref{tab:auc_changing_seed}).

We also assess whether our graduation predictions reflect other aspects of student success. Specifically, we find a modest correlation (0.34) between our baseline graduation predictions and student GPA, with correlations ranging from 0.37 to 0.42 across the different input sets and models (see Figure~\ref{fig:prediction_correlation_GPA}.A). Notably, the transformer model and logistic regression perform well in this regard. Similar correlation patterns are observed for students admitted based on human rankings, either as direct predictions or ranked within study programs, though these correlations tend to be lower overall (see Figures~\ref{fig:prediction_correlation_GPA}.B and \ref{fig:prediction_correlation_GPA}.C). As such, our models do not only capture study completion better than current admission criteria, but also how well the students perform. We also assess cross-program predictions in Section \ref{sec:counterfactual_predictions_method}, with our findings indicating that students are generally better suited to programs within their chosen field, in line with matching effects.

\paragraph{Prediction policy for program admissions.}\label{sec:admission_prediction_policy}

To evaluate the effectiveness of the different ranking methods, we simulate two counterfactual scenarios inspired by a recently enacted 10\% reduction in student intake at Danish universities~\cite{uddannelses_og_forskningsministeriet_udmontning_2024}. These scenarios involve increasing rejection rates across study programs while maintaining the relative sizes of the intakes. To simulate this, we divide the students into 10 deciles within each study program based on rankings generated by our models using the \emph{academic} input set. We then calculate the mean program completion rate for each decile, as shown in Figure~\ref{fig:contract_main}, separately for students admitted through GPA rankings (Panel~B) and human rankings (Panel~C). The 10\% reduction policy corresponds to rejecting the lowest-ranked 10\% of students in each study program, which allows us to evaluate the efficacy of current ranking methods (based on GPA or human assessment) against those produced by ML models, where lower graduation rates are preferred due to these students not being admitted in the counterfactual scenario. Our results show that both traditional and advanced ML models outperform current GPA and human-based rankings in predicting student success within study programs (Panel~B and~C in Figure~\ref{fig:contract_main}), but note that the students evaluated using human assessment opt-in to this assessment method and differ systematically from the overall student population. Specifically, when we apply a contraction by rejecting the lowest 10\% (represented by bin~1), we find a drop in graduation rates. For those admitted through GPA rankings, the drop is at least 9.8~pp (Figure \ref{fig:contract_main}.B), while for those admitted through human rankings, the drop is at least 6.5~pp compared to algorithmic rankings (see Table~\ref{tab:contracted_students} for further details). Based on the bins in Figure~\ref{fig:contract_main}, these findings are robust for up to 20\% of the admitted population, after which performance differences between GPA, human assessment and ML are less clear. 

The best-performing model is the transformer, followed by the LSTM, gradient boosted trees and logistic regression. These findings indicate a trade-off between model complexity and performance, whether measured using AUC or contraction curves. Contraction curves for all input data types are shown in Figure~\ref{fig:contract_appendix}.

If policymakers were to reduce student intake based solely on algorithmic risk scores, without considering specific study program admissions, we can evaluate overall risk by computing the mean predicted completion rates for all students rather than separately by program. This approach is reflected in our ungrouped contraction curves (see Figure \ref{fig:contract_appendix_ungrouped}). The analysis shows a similar clustering of models when using the \emph{academic} input set, but with a lower graduation rate in the lowest bin, which highlights the unequal distribution of students likely to drop out across different study programs and the models' ability to predict this correctly.

We also examine heterogeneity across academic fields (Figure ~\ref{fig:contract_appendix_field_specific}) to assess how well students admitted based on GPA rankings are screened across different disciplines. Due to small sample sizes and privacy concerns, human rankings are excluded from this analysis. We find that the effectiveness of algorithmic risk scores remains relatively consistent across fields. However, the screening ability of GPA varies considerably, e.g., in fields such as \emph{Health and Welfare} or \emph{Education}, GPA proves to be less effective as a screening tool.

\begin{figure}[H]
    \centering
        \centering
        \includegraphics[width=0.8\linewidth]{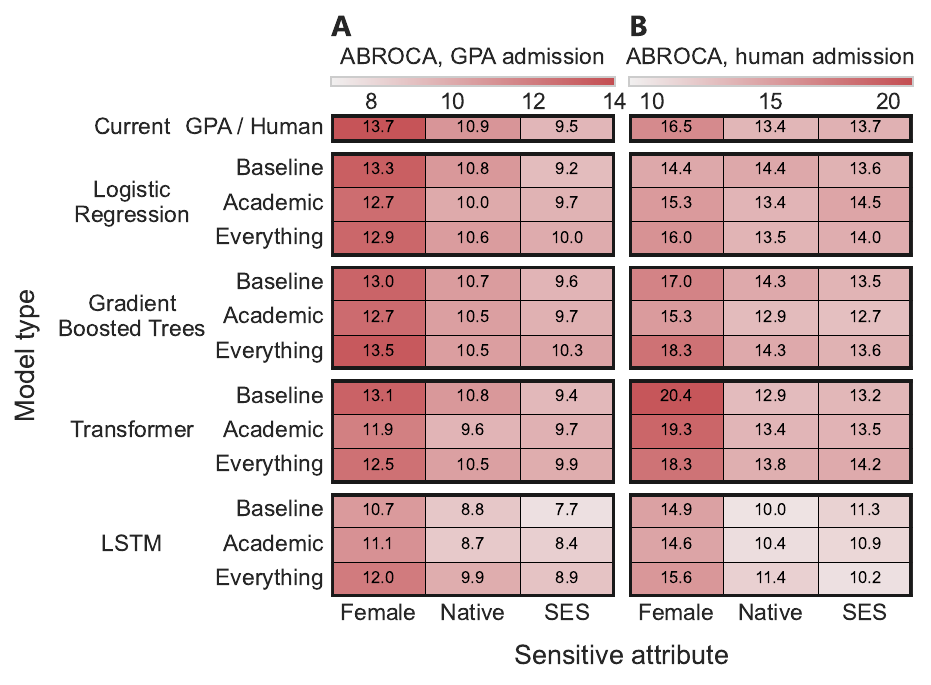}
        \caption[Fairness of admission criteria and algorithms]{\textbf{Fairness of admission criteria and algorithms.} Measures of fairness for different models and inputs based on the ABROCA metric, weighted across institutions. Higher ABROCA values indicate more unequal performance, with 0 corresponding to no performance difference between groups. The fairness scores are divided by admission method: Panel~A for GPA-based admissions and Panel~B for human assessment. The columns display the fairness measure by sensitive attributes: whether a student is a Danish native (\emph{Native}), sex (\emph{Female}), and whether the student is above or below median socioeconomic status (\emph{SES}).}
    \label{fig:abroca}
\end{figure}

\paragraph{Fairness.}\label{sec:fairness_results}
A key aspect of regulating decision algorithms is assessing fairness by evaluating equal treatment across individuals based on sensitive attributes.  
We investigate three attributes that cover different dimensions of human life: socioeconomic status, nativity to country, and biological sex.
To evaluate fairness, we assess whether our models and the admission criteria are equally accurate across these subgroups using the ABROCA (Absolute Between-ROC Area) measure \cite{gardner_evaluating_2019}, a merit-based fairness metric that is independent of arbitrary thresholds, offering a general assessment of fairness (see details in Section~\ref{sec:fairness_methods}) \cite{holmes_algorithmic_2022}. In our setting, the ABROCA also allows us to examine the fairness of current admission criteria. We display this comparison in Figure~\ref{fig:abroca}.
Overall, our models and the admission criteria perform similarly in terms of fairness across these sensitive attributes. One exception is that the LSTM shows better fairness performance regarding sex and socioeconomic status, especially for students admitted through human assessment. In fact, the LSTM demonstrates higher fairness than the currently implemented systems across all dimensions (see standard errors in Figure~\ref{fig:abroca_appendix_standard_errors}). This suggests that it is possible to trade off some performance gains for improved fairness compared to the current policies. ABROCA scores across all input data types are shown in Figure~\ref{fig:abroca_appendix}, where we observe no systematic difference in fairness according to the type of input data used, and that the lower ABROCA scores for LSTM along the dimensions of nativity and socioeconomic status are consistent across input data types. 

Several other fairness metrics, such as sufficiency, independence, and separation (as defined in \cite{barocas_fairness_2023}) are commonly used in algorithmic audits. However, these metrics require a predicted score and thus cannot be calculated for rankings without further information about the underlying generating process. Therefore, we report these metrics for our prediction models only, as described in Section~\ref{sec:fairness_methods} and displayed in Figure~\ref{fig:fairness_appendix}. For all models and sensitive attributes, we find that we cannot reject sufficiency, indicating that the models do not become more accurate when including the attribute. However, we do reject separation and independence for nearly all attributes and models, which means that we observe different error rates across groups (separation) and differing mean predicted scores across groups (independence), reflecting the inherent trade-offs between the fairness criteria \cite{chouldechova_fair_2017, kleinberg_inherent_2017}. The bias in the models generally goes against historically disadvantaged groups, except in the case of sex, where we observe bias against males.

\subsection{Public returns to algorithmic admissions}
A key question for policymakers is how much value a specific public investment would generate. To assess this, we estimate the Marginal Value of Public Funds \cite{hendren_unified_2020} for algorithmic admissions by comparing two counterfactual policies, both rejecting the 10\% lowest-ranked students. The first policy uses the \emph{academic} input data and is simulated from the logistic regression, which has the lowest performance among our models and hence yields a conservative estimate. The second policy serves as the baseline, reflecting the actual admission rankings in use. Detailed descriptions of both policies and our calculations are provided in Section~\ref{sec:marginal_value_of_public_funds}. Consistent with prior research, we assume that increasing the number of graduates produces positive social benefits, as education yields both monetary and non-monetary returns. We estimate that even the worst-performing algorithmic policy would increase government revenue by 86~million USD annually for each cohort (approximately 0.02\% of Denmark's GDP \cite{international_monetary_fund_denmark_2024}) compared to the baseline policy. This additional revenue would come from increased income and consumption taxes. Our best estimate of the costs of implementing these policies includes 1~million USD in initial costs and 16.6~million USD in running costs, with 15.6~million USD of the running costs resulting from reduced screening performance when human judgment overrides the algorithmic decisions.

\begin{figure}[H]
    \centering
        \includegraphics[width=\linewidth]{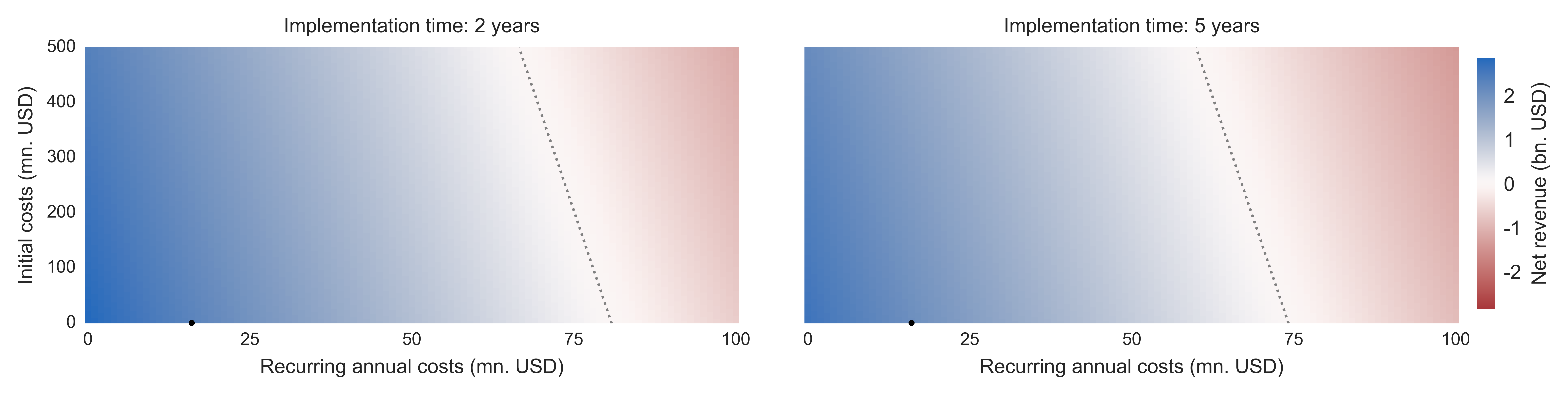}
    \caption[Net government revenue for different scenarios.]{\textbf{Value of adopting prediction-based admission for different scenarios} The figure shows the net present value for various cost and development time scenarios. The two panels depict two different time horizons for implementing the prediction-based admission policies, corresponding to similar delays in revenue generation. We use the lowest revenue estimate if algorithmic screening were adopted, which is based on a logistic regression. The dotted line indicates a revenue of 0, and the black marker indicates our estimated cost.}
    \label{fig:revenue_heatmap_all}
\end{figure}

Given the uncertainty in the cost of implementing algorithmic admissions policies, we estimate the net government revenue under various cost scenarios, as shown in Fig.~\ref{fig:revenue_heatmap_all}. We account for different delays in policy implementation, as well as variations in fixed and variable costs, discounting all values back to year~0. Our analysis shows that switching to algorithmic admissions yields infinite Marginal Values of Public Funds even for initial and running costs significantly higher than our estimates. This result is driven by the scalability of algorithmic policies. Notably, when we focus on implementing the algorithmic ranking only to students admitted through human assessment (representing only 19\% of the sample), we observe a more restrictive set of cost scenarios where policy implementation remains feasible (Fig.~\ref{fig:revenue_heatmap_quota_2}). These findings align with results from prior research \cite{ludwig_unreasonable_2024}, which also emphasizes that scalability is key to making such policies cost-effective.

Lastly, we assess the benefits of increased model complexity. While we observe a 1.1 pp improvement in the AUC metric when using the transformer model compared to the worst-best performing model (logistic regression), it is important to weigh whether this leap in model complexity---from an intrinsically interpretable model to a deep learning method---is justified. The modest performance gain comes with increased challenges in meeting regulatory compliance, a point we explore further in the Discussion section. However, the transformer model generates 136 more graduates compared to logistic regression, translating to an estimated additional 31~million USD in annual gains. This financial benefit could offset even substantial cost increases associated with the added complexity and regulatory demands of using more sophisticated models. 

\section{Discussion}\label{sec:discussion}

Predictive technologies and ML tools hold the potential to enhance college admission policies by improving accuracy and fairness. However, realizing these benefits depends on the careful development and implementation of these tools. In this discussion, we explore the implications of integrating ML models into public policy, focusing on the trade-offs between complexity and explainability, their roles in formalizing and diagnosing within admission policies, and the broader policy implications of adopting such technologies.

One fundamental aspect in deploying predictive models is the extent to which a model can be opaque, or ``black-box", a concern often tied to the accuracy/explainability trade-off \cite{bell_its_2022}. This raises important questions about how \emph{actionable} predictions are if we do not fully understand what drives them \cite{liu_reimagining_2023}, and whether simpler mechanisms might outperform more complex predictive models \cite{perdomo_difficult_2023}. In our context, all risk scores generated by the models are capable of ranking applicants within the admissions system, making them equally actionable, regardless of the complexity of the model. We also find that more complex models can yield better performance and/or fairness metrics, but that the largest performance increase stems from using \emph{any} machine learning model rather than the currently implemented policies. At the same time, transparent and explainable models are generally favored because they enhance public and expert trust, accountability, transparency, and ease of use and contestation \cite{veale_fairness_2018, kolkman_how_2016, rudin_stop_2019}.

\paragraph{Regulatory compliance.}\label{sec:regulatory_compliance} According to the recently enacted EU AI Act \cite{european_parliament_artificial_2024}, ML systems used in college admissions would be automatically be classified as high-risk applications. This classification imposes strict compliance requirements \emph{``as regards risk management, the quality and relevance of data sets used, technical documentation and record-keeping, transparency and the provision of information to deployers, human oversight, and robustness, accuracy and cybersecurity}'' \cite{european_parliament_artificial_2024}. Although the detailed implementation of these regulations is still pending, we offer a discussion of how some of these requirements would apply to algorithmic solutions for student admissions.

Our results demonstrate that algorithmic admission policies outperform both GPA-based and human evaluations, with even the least effective algorithm showing significant long-term economic benefits (see Section~\ref{sec:marginal_value_of_public_funds}). A major factor contributing to this success is the availability of centralized registry data in Denmark, which covers the entire country and provides an exceptionally clean, complete, and well-documented dataset. This high-quality data ensures the reliability and accuracy of the predictions generated by the algorithms, which would help meet the stringent data quality requirements under the EU AI Act. However, these conditions may not be easily replicated in other countries or contexts where such comprehensive datasets are unavailable.

In terms of transparency, the worst-performing model wrt.~screening, logistic regression, is inherently interpretable, offering clear insights into how predictions are made. By contrast, the best-performing model, the transformer, can only provide plausible explanations using current saliency-based methods, see \ref{fig:explainability_saliency} for an example. However, there are significant caveats to the explanations of the deep learning models. Interpreting decisions made by deep learning models, such as transformers, remains a fundamentally unsolved challenge. State-of-the-art explainability methods can produce inconsistent results and may not accurately reflect the model's true decision-making process \cite{adebayo_sanity_2018, atanasova_faithfulness_2023, jacovi_towards_2020}. 
These inconsistencies are often addressed using ad-hoc heuristics, which can mislead stakeholders and end-users \cite{krishna_disagreement_2024}. The lack of a widely accepted definition of a valid algorithmic explanation in policy further complicates the issue \cite{nannini_explainability_2023}, and thus it remains unclear when a model is considered sufficiently transparent. Some advocate for inherently interpretable models, such as logistic regression, that offer transparency by design \cite{rudin_stop_2019}, while others argue that post-hoc explanations, which attempt to clarify how complex models reach their decisions, are sufficient \cite{bell_its_2022}. Therefore, if transparency and the right to explanation are prerequisites in high-stake applications like college admissions, the deployment of deep learning models, such as transformers, hinges on both development of reliable and faithful explainability methods and accepting the usage of models which can only be interpreted post-hoc. Moreover, clear guidelines on what constitutes a valid explanation are necessary to ensure that these methods genuinely assist end-users.

Even if a model is sufficiently interpretable for actionable human oversight, allocation problems like student admissions present additional challenges. Overriding one prediction can affect the rankings of other applicants, making decisions interdependent. This interdependence demands careful consideration to ensure effective oversight and the ability to ``override or disregard the system'' when necessary \cite{edwards_eu_2021}.

As we observe study completion or dropout for those admitted using the current admission policies, we only selectively observe labels \cite{ludwig_fragile_2021, kleinberg_human_2018}, which is not an issue for our evaluation due to two reasons; (1) the counterfactual policy to which we compare our machine learning methods is a policy which reduces the number of students admitted, and (2) we observe the full rankings of the currently admitted student population, either through GPA or human evaluation. Thus the question is which part of the currently admitted student population not to admit; for these, we observe outcomes and know which part of the current student population would not have been admitted using current policies through the known rankings. We are, however, unable to say anything about how our performance extends to the wider unadmitted population of prospective students. At the same time, model performance varies across different fields of study. Assessing the robustness of these algorithms in terms of performance and fairness for new cohorts, different study programs, and the broader student body---including those who were not admitted---is essential due to the selective labels problem. Such evaluations are critical before considering deployment. In Table~\ref{tab:auc_changing_seed}, we assess the robustness of the models to different initializations of the model parameters, showing that logistic regression exhibits minimal variability in out-of-sample performance compared to more complex models. Additionally, applying these models in other situations would require re-assessing their performance to ensure they remain effective. 

\paragraph{Challenges and opportunities}
Integrating predictive models into admission systems offers both opportunities and challenges. A key concern is a potential for feedback loops affecting student behavior. Unlike the transparent ranking by high school GPA, predictive models may require applicants to spend more time determining which programs are feasible, which could increase the risk of misinformed choices \cite{chrisander_why_2023}. However, advancements in smart matching platforms can alleviate this concern by conveying which programs are likely to be feasible \cite{arteaga_smart_2022}. While we find that ranking students based on risk scores enhances accuracy, it also introduces the risk of manipulation, as applicants may try to game the system by adjusting their provided information. This trade-off mirrors broader concerns in market design, such as the strategy-proof Danish admission system, which intentionally excludes application information to prevent manipulation \cite{roth_economics_1982, bjerre-nielsen_voluntary_2022}. Algorithmic rankings could undermine this policy, encouraging applicants to alter their course choices based on perceived effects on predicted scores, potentially compromising both fairness and accuracy \cite{bjerre-nielsen_playing_2023}.

The applications of predictive models and ML tools in admissions extend beyond simply ranking applicants, both in pre- and post-admission contexts. A computational evaluation of candidates could be used in tools for various purposes, such as replicating human rankings to enhance evaluation efficiency \cite{waters_grade_2014}, providing decision-makers with summary information \cite{liu_reimagining_2023} or inform students about program compatibility, guide scholarship distribution, and improve student tracking \cite{aulck_increasing_2020,bergman_seven-college_2021}. These algorithm-generated predictive rankings could improve applicant screening by compensating for human limitations in processing large amounts of information \cite{kolkman_usefulness_2020, harris_decision-making_2024}. As an example, the current secondary quota system in Denmark tends to select high-performing students through self-selection rather than an in-depth human screening process \cite{gandil_college_2022}. However, integrating algorithmic summaries into this process remains challenging because of human tendencies to ignore or override algorithmic suggestions, which can sometimes result in suboptimal outcomes \cite{angelova_algorithmic_2023}, although there are instances where human judgment may outperform model predictions \cite{angelova_algorithmic_2023, kleinberg_human_2018, ludwig_fragile_2021}. Predictive models could also support early warning systems aimed at improving student outcomes, such as identifying students at risk of dropping out and providing timely interventions \cite{balfanz_chapter_2019}. However, the effectiveness of these systems, alongside issues of access and the potential societal implications for at-risk students, is still being studied \cite{liu_reimagining_2023, selbst_fairness_2019}.

An alternative use of algorithmic admission is to equalize access to education. Our findings suggest that ML-based admission policies do not exacerbate unfairness compared to current practices and, in some cases, may even reduce it, as observed with the LSTM model. Specifically, our models meet the sufficiency criterion, ensuring predictions are well-calibrated across different groups. However, due to inherent trade-offs in fairness measures, they violate the separation criterion \cite{chouldechova_fair_2017, kleinberg_inherent_2017}. To further promote fairness and equalize opportunities for students from minority or low socioeconomic backgrounds, one approach would be to post-process the models to mitigate these disparities. This could involve formalizing affirmative action by adjusting the optimization process to prioritize predictive equality, effectively equalizing outcomes across groups \cite{hardt_equality_2016}. 
\section{Conclusion}

This paper examines algorithmic student admissions by comparing various machine learning models for predicting study completion. We introduce a novel approach to predict student dropout, which relies on representing data in a sequential format and using deep learning architectures such as the transformer. Across our experiments, all machine learning-based approaches outperform current ranking methods used in Denmark, with the largest gain coming from using any machine learning model, even just a simple logistic regression, although the more advanced transformer does offer some additional performance at the cost of interpretability. 

Given the high-risk nature of using machine learning in college admission policies, we conducted a preliminary assessment of its potential benefits and challenges with regard to regulatory compliance. Under a policy that rejects the bottom 10\% of applicants in each program, machine learning models identify subgroups with 9.4 to 12.7 percentage-point lower completion rates than the current policy. Using logistic regression, the worst performing model, would significantly improve the selection of students to retain, and this improvement is economically significant, with an estimated increase of 86~million USD in yearly government revenue. We also believe that the logistic regression could be made in a regulatory-compliant manner due to its inherent interpretability, whereas it is unclear whether transformers can be regulatory-compliant due to fundamental issues in understanding what drives the decisions in the transformer. This suggests that alternative selection strategies than the currently implemented GPA and human screening methods could help retain more graduates, and that these selection strategies need not necessarily be based on deep learning models.

A critical consideration when applying predictive models in admissions is fairness. Our study demonstrates that the models we develop satisfy sufficiency \cite{barocas_fairness_2023} across a wide range of sensitive attributes. While some may advocate for models that satisfy equal opportunity, we use the more comprehensive fairness metric, Absolute Between ROC Area \cite{gardner_evaluating_2019}. This metric allows us to compare our counterfactual data-driven rankings to current admissions practices without relying on arbitrary decision thresholds. Our results show that students admitted based on grade point averages or human assessment are ranked just as fairly by both our tabular models and the transformer. Additionally, the LSTM model ranks the students even more fairly, while also delivering better performance when using algorithmic risk scores. This suggests that, in terms of fairness and accuracy, algorithmic models can offer substantial improvements over traditional admissions methods.

As outlined in our discussion, this study opens several avenues for further exploration in algorithmic public policy, both in admissions and other areas. One key question is whether the efficacy of machine learning can be extended to other policy-relevant domains. At a broader level, both transformers and LSTMs generate numerical embeddings of student academic trajectories, which may have value as covariates in causal machine learning models. Furthermore, there is a need to explore how fair or unfair current policies are in other domains compared to potential algorithmic alternatives. This invites further research into the broader impact and fairness of machine learning in public policy beyond education.

{\footnotesize
\textbf{Author contributions:} M.L.N.\ \& A.B.-N.\ developed the study concept with input from the remaining authors. M.L.N., J.S.R.-P., \& E.C. structured the data. M.L.N.\ \& J.S.R.-P.\ performed the analysis under the supervision of A.B.-N. M.L.N.\ \& A.B.-N.\ wrote the manuscript with input from the remaining authors.} 

\printbibliography[heading=bibintoc]

\newpage
\FloatBarrier
\begin{appendices}
\clearpage

\section{Material and Methods}

In this section, we begin by describing the data sources being used in the study, the institutional setting of higher education applications in Denmark, and the specifics of our target and features. Next, we provide a more detailed explanation of the sequence creation method and introduce the models we selected. We then describe how we calculate saliency scores, define fairness metrics, examine match effects and, finally, estimate the Marginal Value of Public Goods.

\subsection{Data Sources and Ethical Considerations} \label{sec:data_sources_ethical_considerations}

We consider all students in higher education in Denmark for whom primary and secondary school grades are available between 2006 to 2017. All relevant data are sourced from Statistics Denmark registries, allowing us to link information on variables such as sex, immigration status, socioeconomic status, and more. 

We applied for an Institutional Review Board (IRB) review but the study falls outside the purview of the IRB's jurisdiction. Since the IRB is primarily concerned with ensuring the ethical treatment of human participants in research, our study, which relies on secondary data analysis and involves no direct interaction with human subjects, does not require formal IRB oversight. All analyses are conducted at an aggregate level, with at least five observations for each measurement reported, to protect participants' privacy. 

The study aims to assess how machine learning methods can help identify students at risk of dropping out and to compare the performance of these methods with the current admission system. In doing so, we seek to contribute to educational policy and practice by promoting student success and improving decision-making in higher education. To achieve this, we train our models on data from 2006 to 2016 and test the predictions on data from 2017. This approach serves two purposes: it mimics a policy scenario where only historical data are available and prevents data leakage that could occur due to correlated outcomes in a random train-test split \cite{yu_should_2021}. 

\subsection{Institutional Setting} \label{sec:institutional_setting}

Higher education application and admission in Denmark is handled at the study program level, meaning applicants apply to a specific field of study at a particular institution (e.g., economics at the University of Copenhagen). All applications for higher education in Denmark are processed centrally by the Ministry of Higher Education and Science through a system called \textit{The Coordinated Application} (\textit{Den Koordinerede Tilmelding}). Within this system, applicants rank the study programs they wish to attend, and they are offered enrollment at the highest-ranked study program for which they are eligible, based on the number of seats available and their ranking within each program. 

There are two ranking systems for applicants, and study programs have the flexibility to choose how many students to admit using either system:
\begin{itemize}
    \item GPA-based admission is considered the ``default system'` and is mandatory for all applicants. In this system, referred to as \textit{Quota~1}, applicants are ranked in descending order based on their high school GPA. A student's GPA is calculated from both exit exams and continuous assessments.
    \item Human-based screening is voluntary for students. In this system, referred to as \emph{Quota~2}, students are ranked by humans, typically a faculty member from the institution they applied to. Rankings are legally required to be based on academic and objective criteria, but institutions have the autonomy to set specific criteria, which must be announced at least a year before admission \cite{uddannelses-_og_forskningsministeriet_bekendtgorelse_2022}. These criteria vary across programs and may include factors such as \emph{additional education}, \emph{work experience}, \emph{folk high school attendance}, \emph{stays abroad}, \emph{tests} (administered by the institution), and \emph{motivated applications} \cite{bjerre-nielsen_voluntary_2022}.
\end{itemize}  
Human-based admission was introduced to broaden access to higher education based on merit, allowing students without traditional GPAs---such as those from vocational schools---or those with GPAs below the cutoff to enroll \cite{gandil_college_2022}. Admission criteria for most study programs under this system typically include CVs, grades, and essays \cite{gandil_college_2022}. Students are matched with institutions using a Deferred Acceptance mechanism with Voluntary Information Disclosure, specifically the DAVID-Q mechanism as described by \cite{bjerre-nielsen_voluntary_2022}. This corresponds to each institution having two separate rankings of students: One based on GPA (mandatory) and one based on human assessment (voluntary), which are handled independently in the Deferred Acceptance mechanism \cite{gale_college_1962}. Importantly, the mechanism is \emph{strategy-proof} \cite{abdulkadiroglu_school_2003}, meaning students have no incentive to misrepresent their preferences when ranking programs.  Disclosing voluntary information, such as taking an admission test and performing poorly, does not negatively affect their chances of admission based on GPA. The only cost associated with applying through the human-based system is the additional effort required, such as writing a motivated application. This process is visually represented in Panel~A of Figure~\ref{fig:combined-illustration}.

We have access to student GPAs for the entire population, as well as the rankings of students by humans for the subset of individuals who applied for human-based ranking, both sourced from \textit{The Coordinated Application}. This allows us to evaluate the GPA-based and human-based systems as if they were algorithms. However, it is important to note that we can only compute evaluation metrics that require rankings within individual study programs for human-based assessments, as we do not have access to an overall human ranking of all students.

\subsection{Target}

In our analysis, we use degree completion at a study program, conditionally on being admitted, as our target outcome for two main reasons: (1)~it aligns with one of the primary policy goals of the study programs themselves, as a portion of their funding depends on student graduation rates \cite{uddannelses-_og_forskningsministeriet_tilskud_2023}, and providing education is costly for society; and (2)~in the institutional context of Denmark, where students do not pay for higher education and generally receive financial support from the government while studying, the incentives for voluntarily dropping out of a study program are reduced. 
We acknowledge that our target outcome may not fully capture the perspective of students who find value in their even without completing their education, and that the suitability of a target outcome depends on the stakeholders involved.

\subsection{Features}

The core features included in our models are grade transcripts for middle and high school, which provide detailed contextual information about the grades received, e.g., the course name, course level, type of examination, etc. In Denmark, grades are assigned on a 7-point scale, aligned with the ECTS scale, ranging from A to F \cite{ministry_of_higher_education_and_science_grading_2021}.Course level in Denmark indicates the number of years a course is undertaken and is categorized as C, B, and A, corresponding to workloads of 1, 2, and 3 years, respectively. For example, a student might receive a grade~A in mathematics at level~A in an oral exam. Each grade is also linked to a specific academic year and institution. In primary school, this consists only of an institutional identifier and grade level, while in high school, it includes additional information such as the type of high school (e.g., regular or technical) and study line, which  students choose and which determines the minimum required workloads for different courses. In the previous example, the math grade could have been earned at institution 12345 while enrolled in a study line like \textit{Math-A-Physics-A} at a regular high school. This results in a large number of unique combinations in the data. When focusing on courses, course levels, and examination types alone, we observe 649 unique combinations in our training data (see Table~\ref{tab:categorical} in the Appendix). 

We also incorporate information on student applications. In Denmark, students can rank up to eight study programs, and they are admitted to the highest-ranked study program for which they are eligible \cite{gandil_college_2022}. We observe the number of study programs a student applies to, whether the student opts into human assessment for each program, and the resulting human rankings if they opt in. However, we do \emph{not} have access to the non-grade materials used for human assessments, such as essays or CVs. This results in a smaller feature set for our models compared to the information available to human evaluators.

\subsection{Models}\label{sec:models}

A key challenge in educational data mining has been extracting meaningful signals about a student’s academic potential from noisy data. Most existing studies rely on the common tabular format \cite{xiao_survey_2022}. However, more recent research has explored the use of deep learning models in educational contexts, employing data representations and model architectures tailored to specific problems. For example, some studies have focused on predicting online course completion both before and during course enrollment \cite{kim_gritnet_2018}, while others have investigated methods for measuring students' implicit knowledge \cite{pu_self-attention_2022}.

\subsubsection{Sequence Creation}\label{sec:sequence_creation}
We rely on multiple data sources for each student, including sociodemographic information, grade transcripts, and enrollment records, as shown in Panel~B of Figure~\ref{fig:combined-illustration}. Each data source is preprocessed independently, breaking the task into simpler,  manageable steps while allowing us to use tried-and-true preprocessing techniques. To prepare the data for modeling, we transform the tabular data into sequences of events, arranging them in chronological order. Each event is represented by categorical variables that describe different aspects of that event. To maintain consistency, all continuous numerical variables are converted into categorical variables. This is done by first applying winsorization to the top and bottom 5\% of values to mitigate the influence of outliers, and then converting the remaining values into percentiles.

For a given student, there exist \( C\) sequences, denoted as \( \mathcal{s}^c = \{v_1^c, v_2^c, \ldots, v_n^c\} \), where \( n \) represents the number of events for that student. Each $c$ in \( C\) corresponds to a specific aspect of events, with each aspect corresponding to a single column in the tabular representation. The value $v_i^j$ denotes the \(i\)\textsuperscript{th} value in the \(c\)\textsuperscript{th} sequence. We represent each value as a string, which means that, in practice, each sequence corresponds to a sentence, with each value being a word. The words at position \(i \) encode information about the same event \( i \) across all sentences. We then trim sentences by removing the initial words observed or pad sequences to a fixed length $L$, corresponding to the 95\textsuperscript{th} percentile of sequence lengths. Sociodemographic variables, when included, are never removed during trimming, ensuring their consistent inclusion across sequences. Padding is done using a special token {\tt [PAD]}.

As shown in Panel~C of Figure~\ref{fig:combined-illustration}, many sentences may lack a specific word \( v_i^c \) associated with an event due to certain aspects being irrelevant. For example, if the \(i\)\textsuperscript{th} event is an enrollment, there will be no associated value in the grade aspect, as the grade aspect is irrelevant for enrollment events. In such cases, we replace these words in irrelevant aspects with a null token ({\tt [Null]}). Finally, we prepend a {\tt [Null]} token at the beginning of all sentences and generate a new sentence starting with a single classification token ({\tt [CLS]}) followed by {\tt [Null]} tokens until the sequence reaches the fixed length $L$. The {\tt [CLS]} token is used to represent the full set of sentences and is utilized for making predictions in subsequent tasks, as described in \cite{devlin_bert_2018}. Strings occurring fewer than 250 times in the training set are replaced with a special unknown token {\tt [UNK]}, to manage rare occurrences.

All the unique strings form the vocabulary, denoted as $\mathcal{V}$. To represent these concepts numerically, we convert them into binary vectors using a one-hot encoding scheme. While the vocabulary size $|\mathcal{V}|$ is large, we reduce the dimensionality of one-hot-encoded vectors by embedding them into a lower-dimensional space using a linear mapping, represented as $E_{v_i^c} \equiv f(v_i^c) \in \mathbb{R}^H$, where $H$ is the dimensionality of the embedding space. This linear mapping is optimized jointly with the other model parameters, except for $E(\texttt{[Null]})$, which is specifically set to map to a vector of zeros. 

We now have a matrix of dimensions $L \times C \times D$, which we reduce to a matrix of dimensions $L \times D$ by summing over the aspects for each event: 

 \[ AGG_i =  \sum_{c=1}^C E_{v_i^c}. \] 

\begin{figure}
    \centering
    \begin{adjustbox}{center}
    \includegraphics[width=0.9\textwidth]{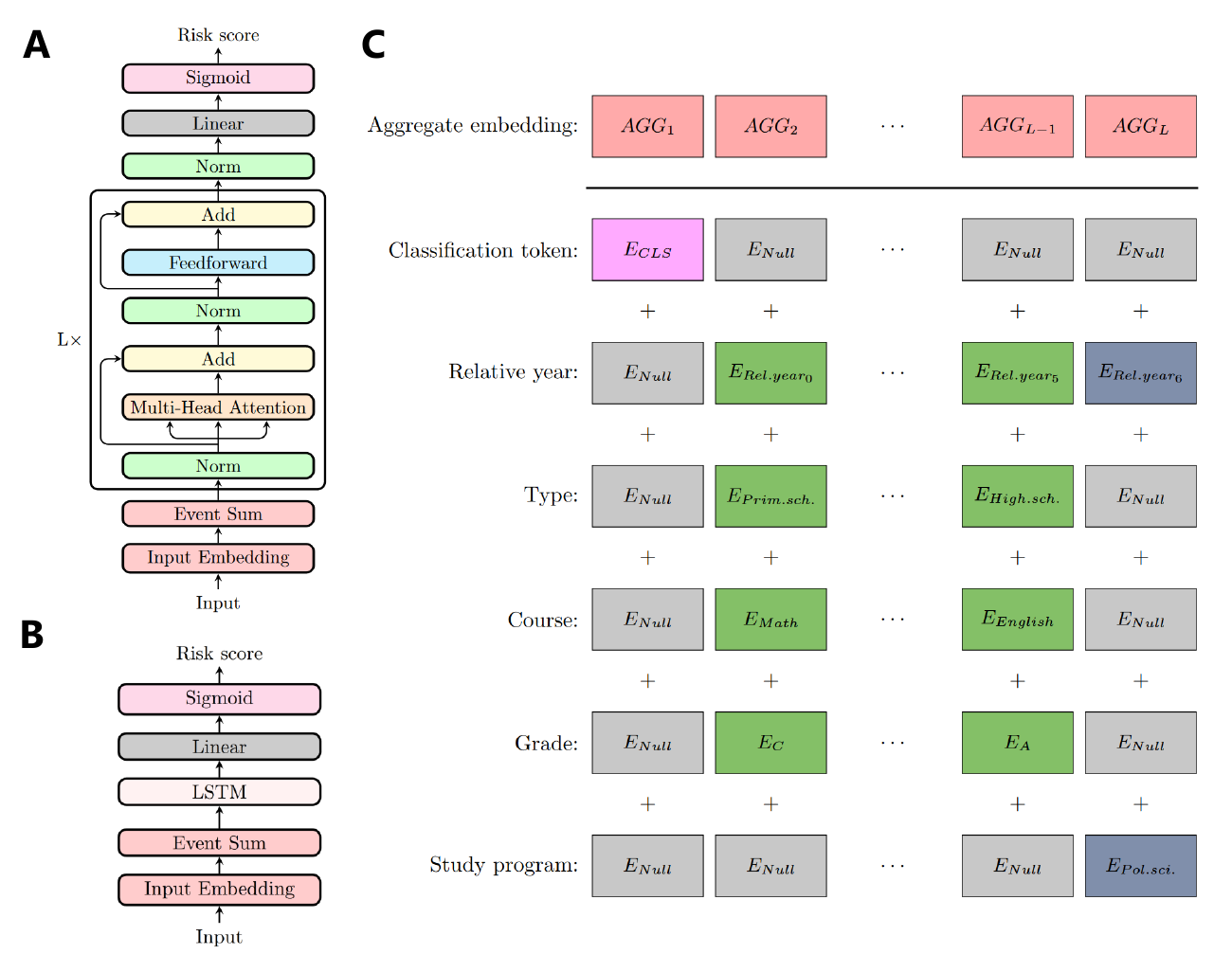}
    \end{adjustbox}
    \caption[Sequential model architectures with aggregate embeddings.]{\textbf{Sequential model architectures with aggregate embeddings} Panel~A illustrates the pre-norm encoder-only transformer architecture, while Panel~B presents the LSTM architecture. Panel~C provides an example of input embedding and summation, corresponding to the red elements shown in Panels~A and~B. For each event, values are first embedded into vectors, and the aggregate embedding of each event is represented as a summation of these vectors. This process matches the input embedding and summation shown at the bottom of Panel~A.}
    \label{fig:transformer_architecture}
\end{figure}

This $(L \times D)$-dimensional vector can be used as input to any deep learning architecture that processes sequences of fixed dimensions, making our representation usable in many different contexts. This approach allows us to create sequences of contextualized events, where multiple columns provide context for each event. The aggregation method is visualized in Panel~C of Figure~\ref{fig:transformer_architecture}.

\subsubsection{Data Input Variants}\label{sec:input_variants}

We develop six distinct models, both sequential and tabular, varying the input data used. We construct a baseline GPA input data set, consisting solely of students' GPA. Next, we construct an \emph{academic} input dataset, designed to mirror the requirement of objective and academic criteria that apply to the human assessment protocol. We then extend this dataset in various ways: The \emph{human assessment} input dataset incorporates human rankings, acknowledging the additional insights these rankings may provide. The \emph{application} input dataset extends the academic input by including comprehensive application details, such as rankings of both enrolled and non-enrolled study programs. The \emph{sociodemographic} model extends the academic input by adding contextual information on the institutional environment (e.g., selection into specific primary and high schools), and sociodemographic information about students and their parents. Lastly, the \emph{everything} model encapsulates all elements from the preceding models, providing a multifaceted approach to predicting study program completion. The specific details of the input variables included in each variant of the sequential models are outlined in Table~\ref{tab:variables_transformer}. In practice, we arrange all events chronologically and include all events occurring in the same year or prior to enrollment, while filtering out irrelevant event types (e.g., excluding application events and sociodemographic events from the \emph{academic} models) and contextual details for each event (e.g., excluding the institutional ID associated with grades in primary and high school from the \textit{academic} models). The variables used in the variants of the tabular models are presented in Table~\ref{tab:variables_baseline}. 

\subsubsection{Sequential Models}\label{sec:sequential_models}

To predict an outcome from these sequences, we use deep learning models suited for classification tasks, with several options available in natural language processing. For our sequential models, we use an encoder-only transformer \cite{devlin_bert_2018} and an LSTM (Long Short-Term Memory) model \cite{hochreiter_long_1997}. The attention mechanism in the transformer allows the model to capture long-range dependencies in the data \cite{vaswani_attention_2017}, while the LSTMs have been widely used in the educational context \cite{dalipi_mooc_2018} and are common in natural language processing, although transformers are increasingly favored. Given the widespread use of these architectures and their standard implementations, we do not provide a detailed description here. Instead, we refer readers to \cite{vaswani_attention_2017, devlin_bert_2018, nguyen_transformers_2019} for detailed discussions of the transformer architecture and to the PyTorch documentation for the LSTM. Both the transformer and LSTM modules are implemented in PyTorch. Visual illustrations of the transformer and LSTM architecture can be found in Panels~A and~B of Figure~\ref{fig:transformer_architecture}. Model information can be found in Table \ref{tab:transformer_model_info}.

Model optimization is conducted using AdamW \cite{loshchilov_decoupled_2018} with the hyperparameters $(\beta_1, \beta_2, \lambda) = (0.9, 0.999, 0.01)$, incorporating 100 warmup steps and a Cosine learning rate scheduler. We train using Distributed Data Parallel (DDP) in \texttt{pytorch lightning} \cite{falcon_pytorch_2019}. All models are trained for 10 epochs on four NVIDIA A100 GPUs, minimizing binary cross-entropy loss. Early stopping is applied, monitoring validation loss with a patience of 3 epochs. We do not pretrain our models, focusing exclusively on the downstream task of predicting study program completion. For model selection, we reserve 5\% of the training sample, stratified by year, as validation data. 

\paragraph{Transformers.}

We follow the notation from \cite{devlin_bert_2018}, where $L$ represents the number of layers, $H$ the hidden size, and $A$ the number of self-attention heads, while the feed-forward size is fixed at $4H$. Our models are trained with $L=8$, $H=512$, and $A=8$. Notably, scaling these hyperparameters up or down by a factor of two results in similar performance on the validation set, indicating robustness to hyperparameter variations. A dropout rate of 10\% is applied to all non-embedding layers. For predictions, we extract the \texttt{[CLS]} token from the final layer and use a simple linear layer for classification. The GELU activation function is employed, and normalization is performed before the attention mechanism \cite{nguyen_transformers_2019}. The models are trained with a learning rate of $5\cdot10^{-4}$ and a batch size of 512 (equivalent to an effective batch size of 2048). Exploration within a learning rate range of $1\cdot10^{-4}$ to $1\cdot10^{-3}$ resulted in comparable performance on the validation set.

\paragraph{LSTM.}

We train LSTM models with a hidden size and input size of 128, using 2 layers and applying a dropout rate of 20\%. Exploration across a range of hidden and input sizes (64 to 512), layers (1 to 3), and dropout rates (10\% to 40\%) yielded similar performance on the validation set. For predictions, we extract the last hidden state and pass it through a simple linear layer for classification. Training is performed with a learning rate of $5\cdot10^{-4}$ and a batch size of 128 (equivalent to an effective batch size of 512). Exploration with higher learning rates (up to $1\cdot10^{-3}$) and different batch sizes yielded comparable results, while the lower learning rates resulted in inferior performance.

\subsubsection{Baseline Models}

For our baseline tabular models, we use logistic regression and gradient-boosted trees, specifically XGBoost \cite{chen_xgboost_2016}. Grade information is encoded by including the GPA and representing grades through the mean grade within each course. Additionally, we calculate and include the mean, standard deviation, and count of grades within three broad fields: \emph{STEM}, \emph{Languages}, and \emph{Other}. Aggregate information is computed separately for primary and secondary schools (see summary statistics in Table~\ref{tab:course_aggregates}). Missing values are handled differently across models: in logistic regression, we apply median imputation, while XGBoost handles missing values internally. In unknown values are encountered during one-hot encoding in the test year, a vector of zeros is used. 

We select the best hyperparameters through 3-fold cross-validation coupled with a randomized grid search, selecting the hyperparameters that yield the highest average out-of-sample AUC score across the left-out folds. We evaluate 30 hyperparameter combinations for both gradient-boosted trees and logistic regression, drawing from the distributions specified in Table~\ref{tab:baseline_hyperparams}.

\subsection{Explainability}\label{sec:explainability_methods}

We use saliency scores to examine how the sequential models generate predictions. These scores are computed using the \texttt{InputXGrad} method \cite{shrikumar_learning_2019} with $l2$-norm aggregation, as this approach has been shown to most closely mimic human annotation across various Natural Language Processing tasks \cite{atanasova_diagnostic_2020}. 

The saliency score $S_j$ for each numeric input $x_j$ is calculated as the partial derivative of the neural network ($\texttt{NN}(\cdot)$) with respect to the input, multiplied by the input:
$$S_j \equiv x_j \cdot \frac{\partial \texttt{NN}(x_j)}{\partial x_j}.$$
Since each input $v_i^c$ maps to a high-dimensional vector $E_{v_i^c} \in \mathbb{R}^H$, we compute $H$ saliency scores for each input. To aggregate these into a single saliency score for a given input, we use the $l2$-norm \cite{atanasova_diagnostic_2020}. Denoting $S_{jh}$ as the $h$-th element of $S_j$, the saliency for an input $v_i^c$ is given by:
$$\texttt{Attr}_{ic} \equiv \sqrt{ \sum_{h=0}^H S_{jh}^2}.$$
We further define the saliency score of an event~$i$ as the sum of the saliency scores for each aspect of the event:
$$\texttt{Attr}_{i} \equiv \sum_{ c = 1}^C \texttt{Attr}_{ic}.$$

An alternative approach would be to use local surrogate models with tabular input to explain predictions in a human-interpretable manne. We opted not to use this method, as one of the main differences between the models is the inclusion of highly granular data, which cannot be effectively captured or explained by local surrogate models based on tabular representations.

\subsection{Fairness}\label{sec:fairness_methods}

To assess the fairness of our models, we rely on three key concepts: independence, separation, and sufficiency, as described by \cite{barocas_fairness_2023}. These concepts are evaluated in terms of the sensitive attribute ($A$), the target variable ($Y$), and the model's score ($R$), respectively. However, a limitation of these fairness metrics is that they require a score for each observation, whereas the admissions system only provides rankings. To address this, we also use the Absolute Between-ROC Area (ABROCA) metric introduced by \cite{gardner_evaluating_2019}, which allows for a more direct comparison between our models and the current admissions system. These operationalizations of fairness have been applied in previous work on educational data mining \cite{kung_interpretable_2020, holmes_algorithmic_2022}. In this section, we briefly describe each of the fairness concepts and their normative underpinnings. 

\textbf{Independence}, also referred to as demographic parity, requires that the model's score $R$ is independent of the sensitive attribute $A$. The normative assumption underlying this concept is that our model should predict the same graduation rates across various groups defined by the sensitive attribute. To test for demographic parity, we test the null hypothesis $H_0: P(\bar Y |A = a) = P(\bar Y | A = b)$, where $Y$ represents the target outcome (graduation). 

\textbf{Separation}, also known as error rate parity or equalized odds \cite{hardt_equality_2016}, requires that $R$ is independent of $A$, conditional on the outcome $Y$. The normative assumption here is that the model should have the same true and false positive rates across the groups defined by the sensitive attribute, allowing distinctions between groups only when justified by the outcome. To test for equal true and false positive error rates, we test the null hypotheses $H_0: P(\bar Y | Y = 1, A = a) = P(\bar Y | Y = 1, A = b)$ and $H_0: P(\bar Y | Y = 0, A = a) = P(\bar Y | Y = 0, A = b)$.

\textbf{Sufficiency} requires that $Y$ is independent of $A$ given $R$. The normative assumption is that our model should fully capture the information in the sensitive attribute. To test this, we group observations into five bins based on the quintiles of the scores, then test the null hypothesis $H_0: P(\bar Y |A = a, R = r) = P(\bar Y | A = b, R = r)$ for each bin $r$. If any test is rejected, the model fails to satisfy sufficiency.

To perform these hypothesis tests for independence, separation, and sufficiency, we use two-proportion $z$-tests at a 5\% significance level. For sufficiency, given that it requires five tests for each sensitive attribute, we apply a Bonferroni correction to control the family-wise error rate at 5\%, separately for each feature for each model. All other tests are performed without corrections.

\textbf{ABROCA} scores \cite{gardner_evaluating_2019} measure the absolute difference in the area under the ROC curves for the baseline group and the sensitive group. It is defined as:
$$ABROCA \equiv \int_{0}^{1} |ROC_{A = 0}(t) - ROC_{A=1}(t)| dt,$$
where $t$ represents the threshold. 
Since ROC curves evaluate the trade-off between true and false positive rates across groups, this ABROCA metric is conceptually similar to separation and shares its normative foundation: it allows the model to distinguish between groups if such distinctions are justified by the outcome, making it a merit-based fairness metric \cite{holmes_algorithmic_2022}. Smaller ABROCA scores indicate less disparity in model performance between groups, and thus more equitable predictions, with the minimum score being 0. Given that we only observe rankings for each study program and admission system, we compute study program-specific ABROCA scores. To consolidate these into a single metric, we compute a weighted average of ABROCA scores across all study programs within each admission system, using the number of admitted students in each program as weights.

 For our fairness analysis we use three indicators: 1) sex as recorded by Statistics Denmark, 2) whether a student is a first or second generation immigrant, and 3) whether a student is above or below median socio-economic status based on the first principal component of a principal component analysis of their mothers' and fathers' income, wealth and education length.

\subsection{Counterfactual predictions}\label{sec:counterfactual_predictions_method}
In this section, we describe how we examine whether our predictions accurately capture the match quality between students’
backgrounds and their study programs and the calculations underlying Figure~\ref{fig:prediction_decompose_student}.

In Panel~A, we calculate the correlation between the predicted likelihood of completion at the study program of admission and the counterfactual likelihoods of completion for a group of study programs in a given field. Let $p_{i,l,m}$ denote the predicted likelihood of completion of student~$i$ at their observed study program~$l$ in field~$m$, and let $p_{i,j,k}$ denote her counterfactual predicted completion likelihood at study program~$j$ in field $k$.
The data consists of pairs $(p_{i,l,m},p_{i,j,k})$ for all $l\neq j$ and $m \neq k$, from which we calculate the correlation coefficients. 

 We furthermore decompose our predictions into three components: (i)~program-specific completion rates that reflect, e.g., low predicted completion rates in natural science programs (Panel~B); (ii)~the individual student component, which captures, e.g., the high predicted completion rates for students from natural or social sciences backgrounds in other programs (Panel~C); (iii)~the residual component, which accounts for remaining factors, such as how students from services are less likely to complete natural science programs but more likely to complete Health programs (Panel~D). We also add a same field indicator in Panel~E, allowing to assess match effects within fields, finding an estimate of 0.6\% (std. error of 0.02).  

In panels~B through E, we run the following regressions, respectively:
\begin{align}
    p_{i,j,k} & = \alpha + \gamma_i + \epsilon_{i,j,k}, \\ \label{eq:person_fe}
    p_{i,j,k} & = \alpha + \delta_j + \epsilon_{i,j,k}, \\
    p_{i,j,k} & = \alpha + \gamma_i + \delta_j + \epsilon_{i,j,k}, \\
    p_{i,j,k} & = \alpha + \beta \cdot \text{same\_field}_{i,j,k} + \gamma_i + \delta_j + \epsilon_{i,j,k},
\end{align}
where $p_{i,j,k}$ is student~$i$'s predicted likelihood of completion at study program $j$ in field $k$, $\gamma_i$ and $\delta_j$ are individual and study program fixed effects, and $\text{same\_field}_{i,j,k}$ is an indicator for whether study program~$j$ is in the same field as the student's observed study program~$l$.
For each of these equations, we compute the residuals $\hat{\epsilon}_{i,j,k}$. A larger residual corresponds to a higher predicted likelihood of completion than that explained by the model's fixed effects. For example, in Equation~\eqref{eq:person_fe}, a positive residual means that the model predicts a higher likelihood of completion in the considered study program than expected based solely on the students' innate characteristics. The residuals are grouped by the observed field of admission $l$ and counterfactual field of admission $k$, and the mean residuals are reported.

\subsection{Marginal Value of Public Funds Estimate}\label{sec:marginal_value_of_public_funds}
In this subsection, we explain how we estimate the Marginal Value of Public Funds (MVPF), a key metric in economic analysis used to assess the efficiency and effectiveness of public policies \cite{hendren_unified_2020, hendren_case_2022}. Our analysis takes the recently enacted reduction in undergraduate capacity at Danish universities by 10.2\% as a starting point \cite{uddannelses_og_forskningsministeriet_udmontning_2024}. We evaluate and compare two counterfactual scenarios: one in which student reductions are based on our predictive models, and the other using the the current admissions assessment. The MVPF provides a quantitative measure of the additional value generated by allocating one more unit of public resources to a specific policy intervention or program. The MVPF is calculated as the ratio of public benefits to net government cost \cite{hendren_case_2022}:
\begin{equation}
    MVPF = \frac{\text{Benefits}}{\text{Net Govt Cost}} = \frac{\Delta W}{\Delta E - \Delta C}
\end{equation}
where $\Delta W$ represents the benefits accrued by individuals affected by the policy, $\Delta E$ denotes the government's upfront expenditure, and $\Delta C$ is the long-run reduction in government costs. 

We assume that students derive benefits from graduating, in line with the extensive literature on the returns to education, both in monetary terms \cite{psacharopoulos_returns_2018, lovenheim_chapter_2023, dalskov_store_2009} and non-monetary benefits \cite{oreopoulos_priceless_2011}, We assign students uniform weights, implying a welfare gain ($\Delta W>0$) compared to the counterfactual policy, as fewer students fail to complete their education under our models.

We estimate the yearly long-run reduction in government costs ($\Delta C$) from the government revenue generated by the additional earnings of students who complete their degrees. For the return to higher education in Denmark, we rely on the findings of \cite{dalskov_store_2009}, who uses a propensity score matching method to calculate the returns for various educational levels. These returns reflect the difference between private income and government expenditure on education, compared to no higher education. In Denmark, returns to higher education vary by type, ranging from 2.4~million to 8.9~million DKK (2007 prices) per person. We conservatively use the lowest estimate of 2.4~million DKK. To estimate the increased government revenue from higher income, we calculate the rise in tax revenues from income and consumption taxes. We apply a marginal tax rate of 37.7\%, which is the marginal tax rate for individuals receiving unemployment benefits, serving as a lower bound due to the progressive nature of the Danish tax system \cite{finansministeriet_okonomisk_2021}. Additionally, we apply a consumption tax rate of 23\%, as recommended by the Danish Ministry of Finance \cite{finansministeriet_dokumentationsnotat_2019}. This results in a government revenue increase of 900,000 DKK (2007 prices) from income tax and 550,000 DKK (2007 prices) from consumption tax, for a total of 1.4~million DKK (2007 prices). Adjusted to 2016 prices using the consumer price index from Statistics Denmark's \emph{StatBank} (Table~\emph{PRIS8}), this amounts to 1.6~million DKK or approximately 230,000~USD (using a conversion rate of 1~USD to 7~DKK). For the model with the lowest performance, we observe an increase of 377 graduates (see Table~\ref{tab:contracted_students}), corresponding to a revenue increase of 86~million~USD. 

\begin{table}
    \centering
    \begin{tabular}{lrrrrr}
\hline
       &   Transformer &   Gradient boost. &   Logistic reg. &   LSTM &   GPA / Human \\
\hline
\multicolumn{5}{l}{\textbf{Graduates}} \\
\hspace{1em} GPA   &        1464   &            1570   &          1582   & 1559   &        1923   \\
\hspace{1em} Human &         304   &             320   &           322   &  304   &         358   \\
\hspace{1em} Both  &        1768   &            1890   &          1904   & 1863   &        2281   \\
\multicolumn{5}{l}{\textbf{Reduction in dropout}} \\
\hspace{1em} GPA   &         459   &             353   &           341   &  364   &           0   \\
\hspace{1em} Human &          54   &              38   &            36   &   54   &           0   \\
\hspace{1em} Both  &         513   &             391   &           377   &  418   &           0   \\
\multicolumn{5}{l}{\textbf{Graduation rate}} \\
\hspace{1em} GPA   &          42.2 &              45.2 &            45.6 &   44.9 &          55.4 \\
\hspace{1em} Human &          54.8 &              57.7 &            58   &   54.8 &          64.5 \\
\hspace{1em} Both  &          43.9 &              47   &            47.3 &   46.3 &          56.7 \\
\multicolumn{5}{l}{\textbf{\%-point graduation rate difference}} \\
\hspace{1em} GPA   &          13.2 &              10.2 &             9.8 &   10.5 &           0   \\
\hspace{1em} Human &           9.7 &               6.8 &             6.5 &    9.7 &           0   \\
\hspace{1em} Both  &          12.7 &               9.7 &             9.4 &   10.4 &           0   \\
\hline
\end{tabular}
    \vspace{-0.5cm}
    \caption[Characteristics of contracted students across model and admission types]{\textbf{Characteristics of contracted students across model and admission types.} \emph{GPA}, \emph{Human} and \emph{Both} refer to admission type. There are 3,470 contracted students admitted using GPA and 555 contracted admitted using human rankings, for a combined contraction of 4,025. All models use the \emph{academic} input set.}
    \label{tab:contracted_students}
\end{table}

Estimating the cost ($\Delta E$) associated with implementing and operating a prediction model for use in the centralized admission process is challenging. A possible benchmark comes from the cost of developing, implementing, and operating an algorithmic placement model in U.S.\ colleges, as reported in \cite{bergman_seven-college_2021}. These costs ranged from 196,170 to 268,890~USD across six different colleges (2016 prices). The primary drivers of these costs were substantial fixed costs related to developing the algorithm and data entry. Given that our model builds on readily available data, we expect lower data entry costs. However, complying with regulatory requirements, such as ensuring the ``right to explanation'', could increase costs. For instance, human oversight---needed to override the automated system at times---might worsen screening performance. Evidence from the Danish admissions process suggests that human screening is less effective, leading to poorer outcomes \cite{gandil_college_2022}. Although this may technically reduce government revenue rather than directly increasing costs, we account for this as a cost increase for simplicity. Given the substantial societal benefit of each graduating student, worsened screening performance could contribute to a significant portion of ongoing expenses. This largely depends on how regulatory compliance, particularly human oversight, is implemented. To maintain a conservative estimate, we project that developing and implementing a centralized admission algorithm will cost around 1~million USD, with an additional 1~million USD for annual operational costs. To account for the cost impact of worsened screening performance due to human oversight overriding predictions, we estimate that an 18\% reduction in graduates would occur, based on 18\% of judges overriding predictions in bail decisions \cite{angelova_algorithmic_2023}, where we instead use the baseline human or GPA admission rule as the performance. This is a very uncertain estimate, and further work would be needed to quantify these costs more precisely. This would lead to an additional cost increase of 14.1 million USD for GPA-based screening (341 additional graduates) and 1.5 million USD for human evaluation (36 additional graduates), totalling 15.6 million USD across both admission types, where we estimate the cost as the predicted increase of graduates multiplied by the 18\% overriding probability multiplied by the revenue of 230,000 USD per student.

\begin{figure}[H]
    \centering
        \includegraphics[width=\linewidth]{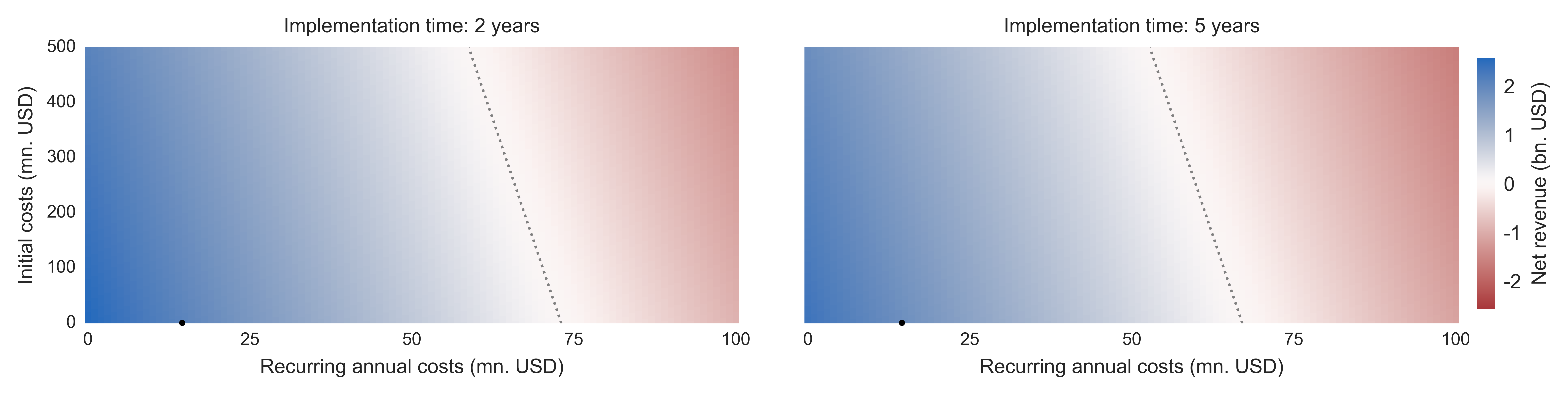}
    \caption[Net government revenue for different cost scenarios for GPA admission]{\textbf{Net government revenue for different cost scenarios for GPA admission.} The figure shows the net present value for various cost and development time scenarios. The two panels depict two different time horizons for implementing the prediction-based admission policies, corresponding to similar delays in revenue generation. We use the lowest revenue estimate if algorithmic screening were adopted, which is based on a logistic regression. The dotted line indicates a revenue of 0, and the black marker indicates our estimated cost.}
    \label{fig:revenue_heatmap_quota_1}
\end{figure}

\begin{figure}[H]
    \centering
        \includegraphics[width=\linewidth]{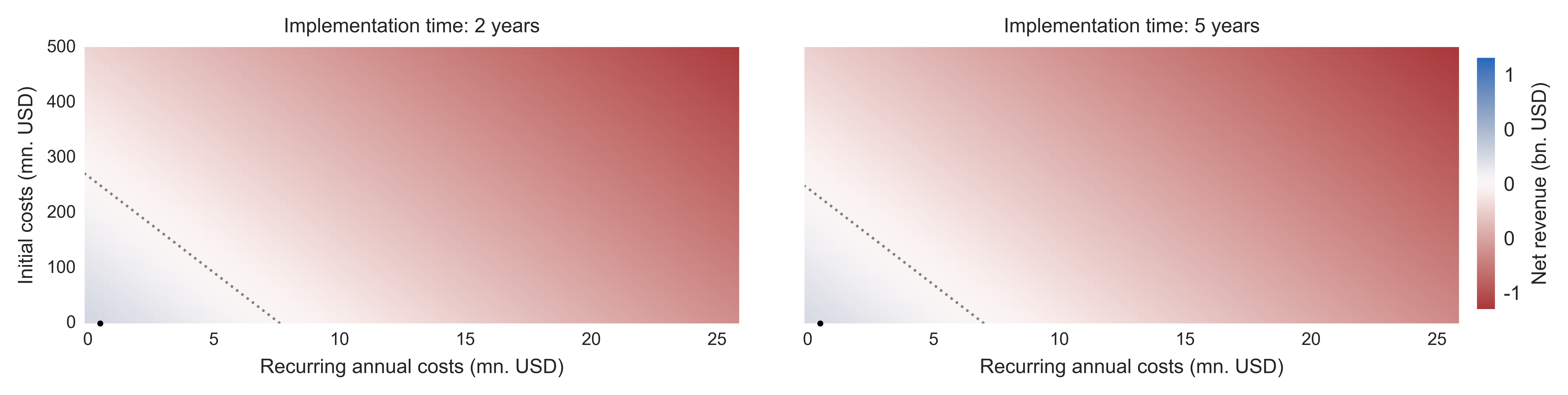}
    \caption[Net government revenue for different cost scenarios for human evaluation admission]{\textbf{Net government revenue for different cost scenarios for human evaluation admission.} The figure shows the net present value for various cost and development time scenarios. The two panels depict two different time horizons for implementing the prediction-based admission policies, corresponding to similar delays in revenue generation. We use the lowest revenue estimate if algorithmic screening were adopted, which is based on alogistic regression. The dotted line indicates a revenue of 0, and the black marker indicates our estimated cost.}
    \label{fig:revenue_heatmap_quota_2}
\end{figure}

Due to the recurring nature of both the revenues and costs, we calculate the net present value (NPV) of the income and costs associated with this policy, as explained in the next paragraph. Given the uncertainty around the cost estimates, we illustrate the net government revenue across different cost scenarios. This is shown for all admitted students in Figure~\ref{fig:revenue_heatmap_all}, for those admitted using GPA in Figure~\ref{fig:revenue_heatmap_quota_1}, and for those admitted via human evaluation in Figure~\ref{fig:revenue_heatmap_quota_2}. Our analysis reveals that net government revenue remains positive over a wide range of cost assumptions, leading to negative government costs and thus infinite Marginal Values of Public Funds.

\paragraph{Net present value calculation.}
To analyze the net present value of a series of revenues (or costs if negative) from year~0 onward under varying conditions, we use a combination of finite and infinite geometric series to discount future payments to their present value. For a finite geometric series  $S_n = ar^0 + ar^1 + \hdots + ar^n$, the present value is calculated as \[ S_n = \sum_{k=0}^n ar^{k} = a \frac{1-r^{n+1}}{1-r} \] where \( a \) is the yearly revenue (or cost), \( r \) is the discount rate, and \( n \) is the number of periods. For an infinite geometric series, $S = ar^0 + ar^1 + ar^2 + \hdots$, the present value is calculated as \[ S = \sum_{k=0}^\infty ar^{k-1} = \frac{a}{1-r} \] where \( a \) and \( r \) are as defined above. It is important to note that these series begin with an undiscounted term, which is not the case for the periods 36-70 and 71+ years. This necessitates an additional discounting of the sum.

We use the official discount rates from the Danish Ministry of Finance, with the discount factor being $R_1 = 1 - 0.035$ for years 1 through 35, $ R_2 = 1 - 0.025$ for years 36 through 70, and $R_3 = 1 - 0.015$ for year 71 and beyond.

For the first period (years 0-35), the present value is calculated using the finite geometric series formula:
\[
\text{PV}_{0-35}(a) = a \frac{1 - (R_1)^{36}}{1 - R_1}.
\]

For the second period (years 36-70), the present value is also calculated using a finite geometric series formula and then discounting it back to year 0:
\[
\text{PV}_{36-70}(a) = a \frac{1 - (R_2)^{35}}{1 - R_2} \times R_2  \times R_1^{35}.
\]

For the third period (years 71 and beyond), the present value is calculated using an infinite geometric series formula, discounted to the present value at year~0:
\[
\text{PV}_{71+}(a) = \frac{a}{1 - R_3} \times R_3  \times (R_2)^{35} \times (R_1)^{35}.
\]

The total present value is the sum of the present values of the three periods:
\[
\text{PV}(a) = \text{PV}_{0-35}(a) + \text{PV}_{36-70}(a) + \text{PV}_{71+}(a).
\]

In some scenarios, revenues may not start in year~0. To account for this, we exclude the revenue for the first $k$ years and calculate the present value for the first period as: 
\[
\text{PV}_{k-35}(a, k) = a \frac{1 - (R_1)^{36 - k}}{1 - R_1} \times (R_1)^k,
\]
where $k$ indicates the year in which revenues start. The total adjusted present value is:
\[
\text{PV}^\text{adjusted}(a, k) = \text{PV}_{k-35}(a, k) + \text{PV}_{36-70}(a) + \text{PV}_{71+}(a)
\]

We decompose revenues into fixed and variable (yearly) revenues (or costs). We assume that a fixed cost occurs in year~0, while variable costs start in year~0 as well. Variable revenues only start after $k$ development years. The net government revenue is calculated as:
\[
\text{Net Govt Cost} = \text{PV}^\text{adjusted}(\text{Rev}, \text{DevTime}) - \text{C}_{Fixed} - \text{PV}(\text{C}_{Var}), 
\]
Where $\text{Rev}$ is the revenue, $\text{C}_{Var}$ is the yearly variable cost, $\text{C}_{Fixed}$ is the fixed costs, and $\text{DevTime}$ is the development time (in years) before revenue starts.
\captionsetup[figure]{list=yes}
\captionsetup[table]{list=yes}

\newpage
\FloatBarrier
\setlength{\cftfignumwidth}{3em}
\setlength{\cfttabnumwidth}{3em}

\renewcommand{\thesection}{SI \arabic{section}}
\renewcommand{\theHsection}{SI \arabic{section}} 
\setcounter{section}{0}   
\renewcommand{\thefigure}{SI \arabic{figure}}
\renewcommand{\theHfigure}{SI \arabic{figure}}
\setcounter{figure}{0}   
\renewcommand{\thetable}{SI \arabic{table}}
\renewcommand{\theHtable}{SI \arabic{table}}
\setcounter{table}{0}   

\section{Supplementary Information}

\renewcommand\cftloftitlefont{\large\textbf}
\renewcommand\cftlottitlefont{\large\textbf}
\renewcommand\listfigurename{List of figures in supplementary information}
\renewcommand\listtablename{List of tables in supplementary information}
\listoffigures
\listoftables

\subsection{Study Program Revenue Calculation}\label{sec:revenue-calculation}

In this section, we briefly describe how we arrive at the estimate of 21,000 to 46,000~USD per bachelor's student. In Denmark, the majority of university funding is allocated through education taximeters, which provide funding based on the number of student full-year equivalents. A full-year equivalent represents the successful completion of exam activities corresponding to one year of standardized study time (60 ECTS credits). Therefore, a three-year bachelor's program corresponds to three rates. Additionally, universities receive a completion bonus for students who complete their studies within a specified timeframe, typically within one year beyond the standard duration for bachelor's degrees and three years for master's degrees. The taximeter rates and completion bonuses vary depending on the type of program, broadly categorized into humanities, mathematics, natural sciences, and others. The 2017 rates are outlined in Table~\ref{tab:takst} from \cite{finansministeriet_forslag_2016}.

These mechanisms lead to payouts ranging from 153,000 to 327,000~DKK per three-year bachelor student who completes their program on time. When converted at an exchange rate of 1~USD to 7~DKK, we arrive at an approximate range of 21,000 to 46,000~USD per student. 

\begin{table}[H]
\centering
\begin{tabular}{lcc}
\toprule
Area & Rate per full-year equivalent & Bachelor completion bonus \\
\midrule
Humanities and others & 44.000 & 21.000 \\
Mathematics and others & 63.200 & 34.100 \\
Natural sciences and others & 92.400 & 49.900 \\
\bottomrule
\end{tabular}
\caption[Rates for full-year equivalents and completion bonus]{\textbf{Rates for full-year equivalents and completion bonus.} Rates are given for our test year of 2017. All amounts are shown in DKK.}
\label{tab:takst}
\end{table}

\FloatBarrier

\subsection{Additional Results}

\begin{table}[H]
    \resizebox{\textwidth}{!}{
\begin{tabular}{lcccccc}
\toprule
Model type & GPA baseline & Academic & Application & Human assessment & Socio-demographic & Everything \\
\midrule
LSTM & 63.16 (0.27) & 67.44 (0.26) & 67.97 (0.26) & 67.47 (0.26) & 67.14 (0.27) & 68.27 (0.26) \\
Gradient boost. & 64.59 (0.27) & 68.45 (0.26) & 69.31 (0.26) & 68.70 (0.26) & 69.02 (0.26) & 69.66 (0.26) \\
Logistic reg. & 64.40 (0.27) & 68.50 (0.26) & 69.29 (0.26) & 68.83 (0.26) & 68.54 (0.26) & 69.44 (0.26) \\
Transformer & 63.76 (0.27) & 69.64 (0.26) & 70.46 (0.26) & 69.79 (0.26) & 68.89 (0.26) & 70.29 (0.26) \\
\bottomrule
\end{tabular}
}
    \vspace{-0.5cm}
    \caption[AUC scores across models]{\textbf{AUC scores across model types and input sets.} \label{tab:auc} AUC scores are calculated on held-out data from the year 2017, with standard errors (in parentheses) calculated using \cite{gildenblat_python_2023}. The features included in each transformer are described in Table~\ref{tab:variables_transformer}, while the features for logistic regression and XGBoost are detailed in Table~\ref{tab:variables_baseline}.}
\end{table}

\begin{table}[H]
    \resizebox{\textwidth}{!}{\begin{tabular}{lcccccc}
\toprule
Model type & GPA baseline & Academic & Application & Human assessment & Socio-demographic & Everything \\
\midrule
LSTM & 63.32 (0.006) & 67.28 (0.081) & 68.11 (0.226) & 67.77 (0.136) & 66.90 (0.095) & 68.30 (0.123) \\
Logistic reg. & 64.41 (0.006) & 68.46 (0.007) & 69.30 (0.004) & 68.82 (0.012) & 68.61 (0.008) & 69.58 (0.004) \\
Gradient boost. & 64.53 (0.046) & 68.74 (0.183) & 69.65 (0.158) & 68.99 (0.184) & 69.30 (0.283) & 70.25 (0.102) \\
Transformer & 63.73 (0.051) & 69.39 (0.051) & 70.69 (0.037) & 69.89 (0.051) & 69.05 (0.086) & 70.37 (0.088) \\
\bottomrule
\end{tabular}
}
    \vspace{-0.5cm}
    \caption[Mean AUC scores across different initializations]{\textbf{Mean AUC scores across different initializations.} \label{tab:auc_changing_seed} Mean AUC scores are calculated on held-out data from the year 2017 for each random seed, with standard errors reported in parentheses. The features included in each transformer are described in Table~\ref{tab:variables_transformer}, while the features for logistic regression and XGBoost are detailed in Table~\ref{tab:variables_baseline}.}
\end{table}

\begin{figure}[H]
    \begin{adjustbox}{center}
    \includegraphics[width=1\linewidth]{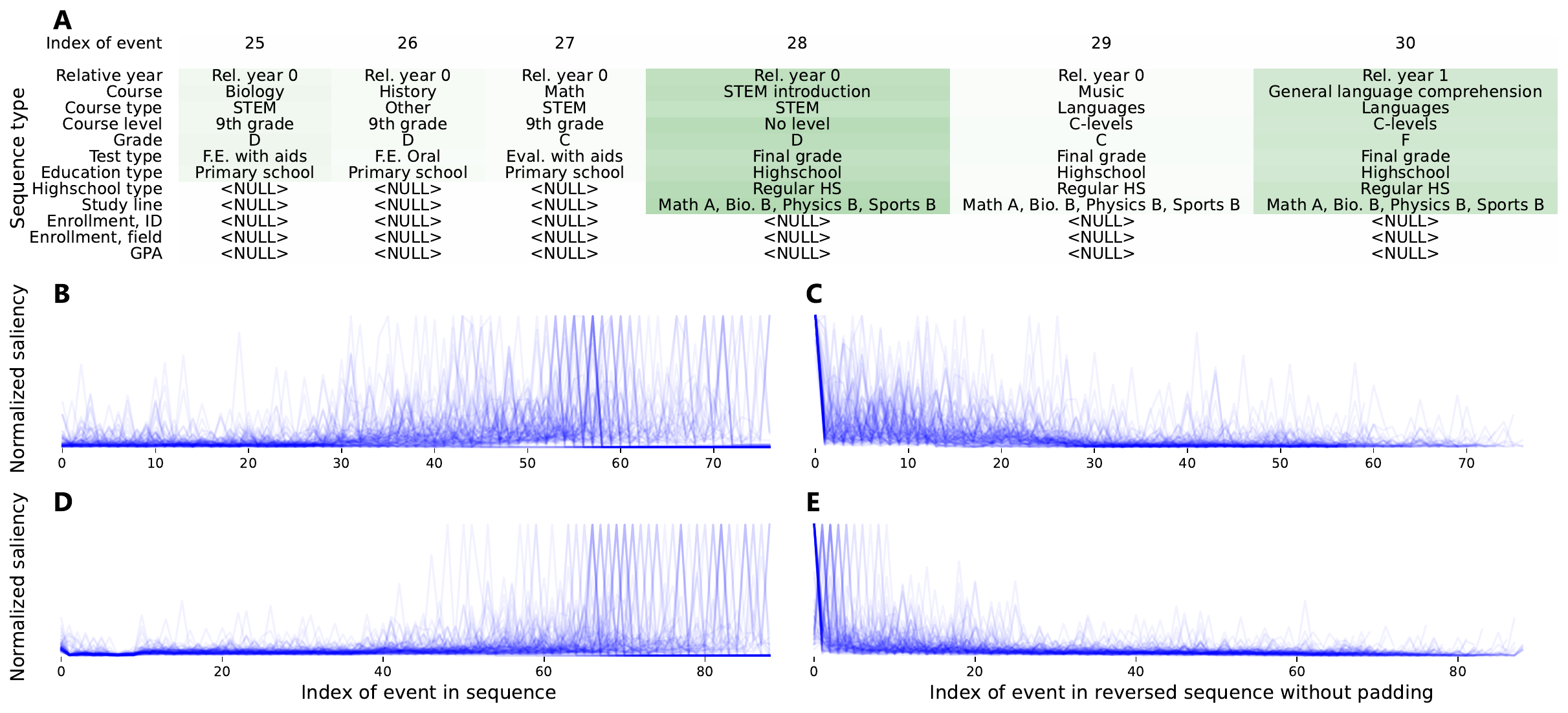}
    \end{adjustbox}
    \caption[Explanations of transformer model]{\textbf{Explanations of transformer model.} Panel~A illustrates the application of saliency scores for local explanations, showing a segment of a sequence representing a student for the \emph{academic} input set. Panels~B and~C display the event saliency across 100 sequences, both in order and reverse order, with padding events removed for the \emph{academic} model. Panels~D and~E present the same analysis for the \emph{everything} input set. Saliency scores are calculated as described in Section~\ref{sec:explainability_methods}.}
    \label{fig:explainability_saliency}
    
\end{figure}

\begin{figure}
    \centering
        \centering
        \includegraphics[width=1\linewidth]{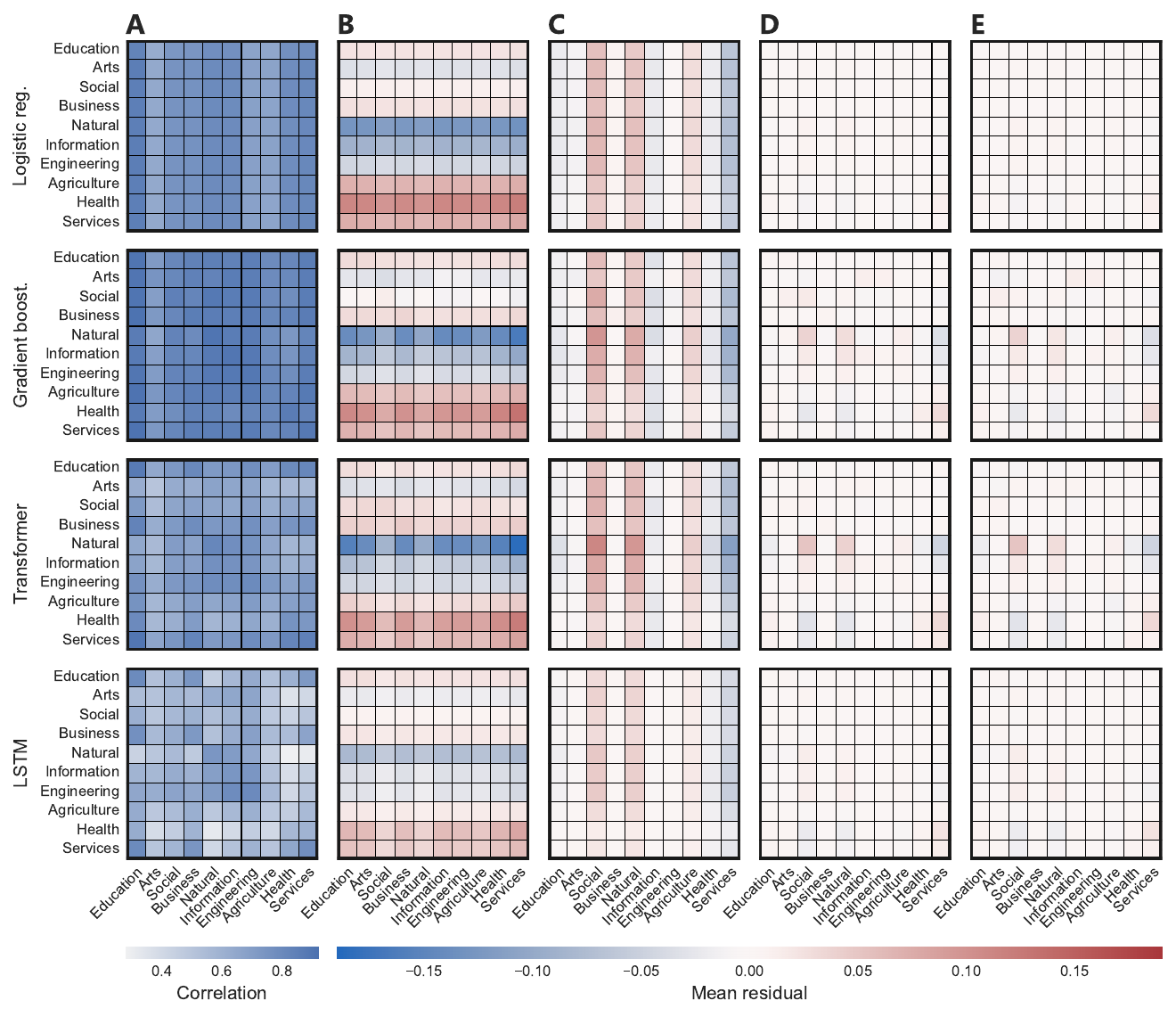}    
    \caption[Correlation between counterfactual predictions and residuals from prediction decompositions]{\textbf{Correlation between counterfactual predictions and residuals from prediction decompositions.} For all heatmaps, the horizontal axis represents the actual field of admission and the vertical axis represents the counterfactual field of admission. Panel~A presents correlations between actual predictions and counterfactual predictions. Panels B-E display residuals from the decomposition of predicted completion scores as a function of covariates. In Panel~B, we control for student ability. In Panel~C, we control for counterfactual study program of admission. In Panel~D, we control for both student ability and counterfactual study program of admission. In Panel~E, we further control for whether the student attended a program in the same field as the counterfactual study program.}
    \label{fig:prediction_decompose_student}
\end{figure}

\begin{figure}
    \centering
        \centering
        \includegraphics[width=.75\linewidth]{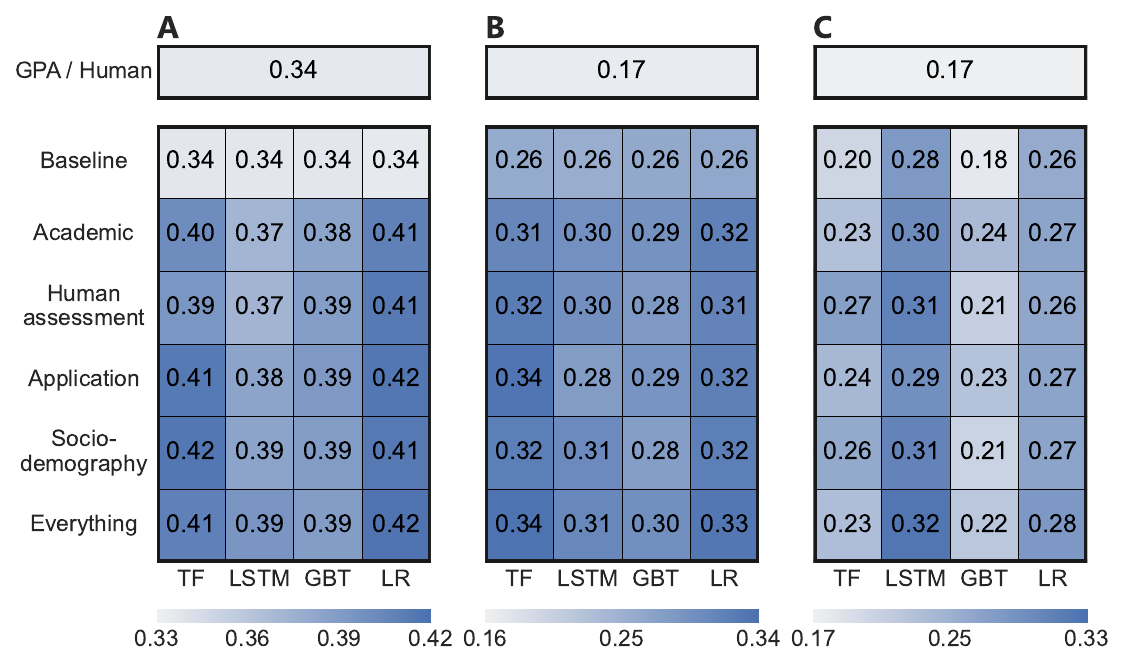}
    \caption[Correlation between predicted completion and first-year GPA]{\textbf{Correlations for predicted completion and first-year GPA.} This figure shows the correlations between the predicted likelihood of completing a program and the actual first-year GPA, computed for students with available first-year grades. Panel~A shows the overall correlations for students admitted via GPA rankings. Panel~B shows the overall correlations for students admitted via human rankings. In both panels, the correlation at the top is between high school GPA and first-year GPA. Panel~C displays the within-study program rank correlations for students admitted through human rankings, with the top correlation representing human ranking. Predicted ranks are computed within each study program for each method, allowing us to compare the model rankings with human rankings, even when overall rankings are not observed.}
    \label{fig:prediction_correlation_GPA}
\end{figure}

\begin{figure}[H]
    \centering
        \includegraphics[width=\linewidth]{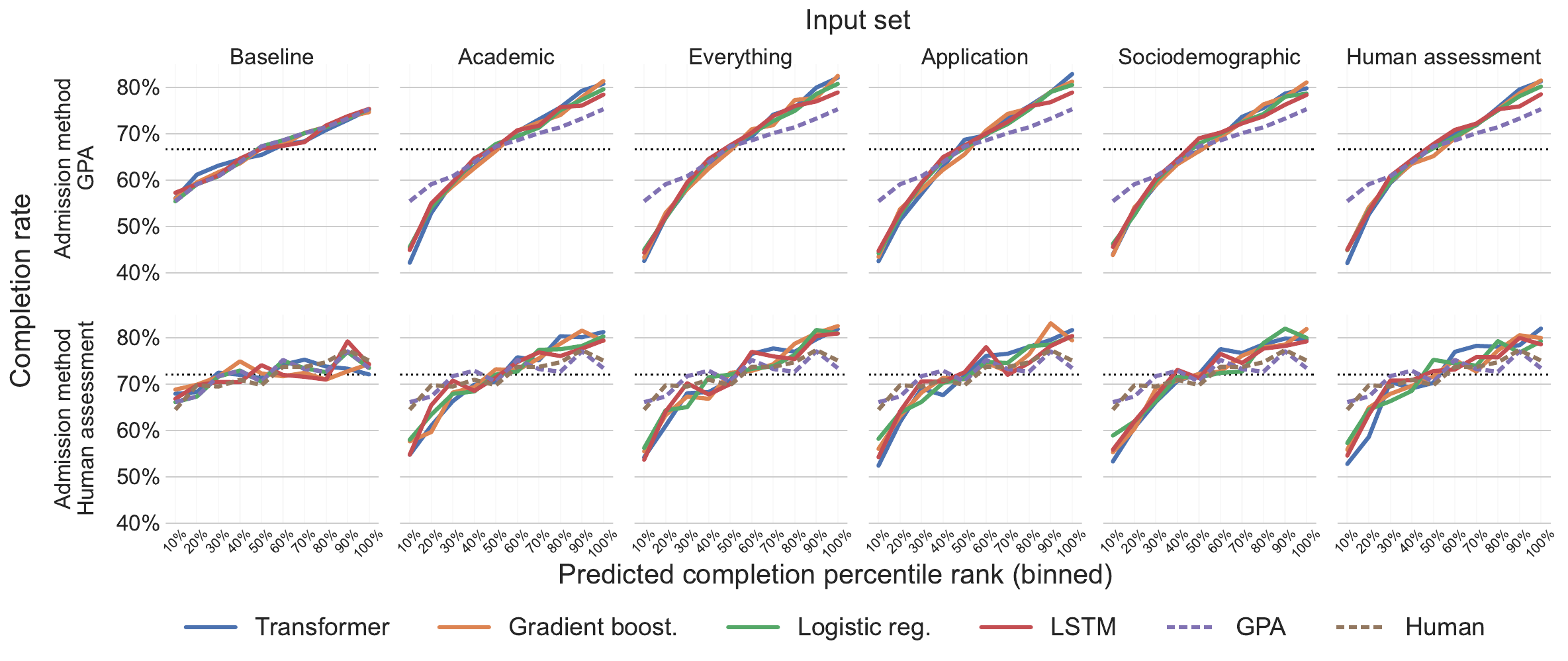}
    \caption[Study program contraction curves across models and student populations]{\textbf{Study program contraction curves across models and student populations.} Students' completion probabilities within each study program are binned in ascending order, and mean completion rates are computed for each bin. The top row presents results for students admitted through GPA-based admissions, while the bottom row displays results for those admitted via human assessment. Perfect ranking would occur if completion rates started at 0\% and then abruptly jumped to 100\%. Dotted lines represent the mean completion rate for each sample.} 
    \label{fig:contract_appendix}
\end{figure}

\begin{figure}[H]
    \centering
        \includegraphics[width=\linewidth]{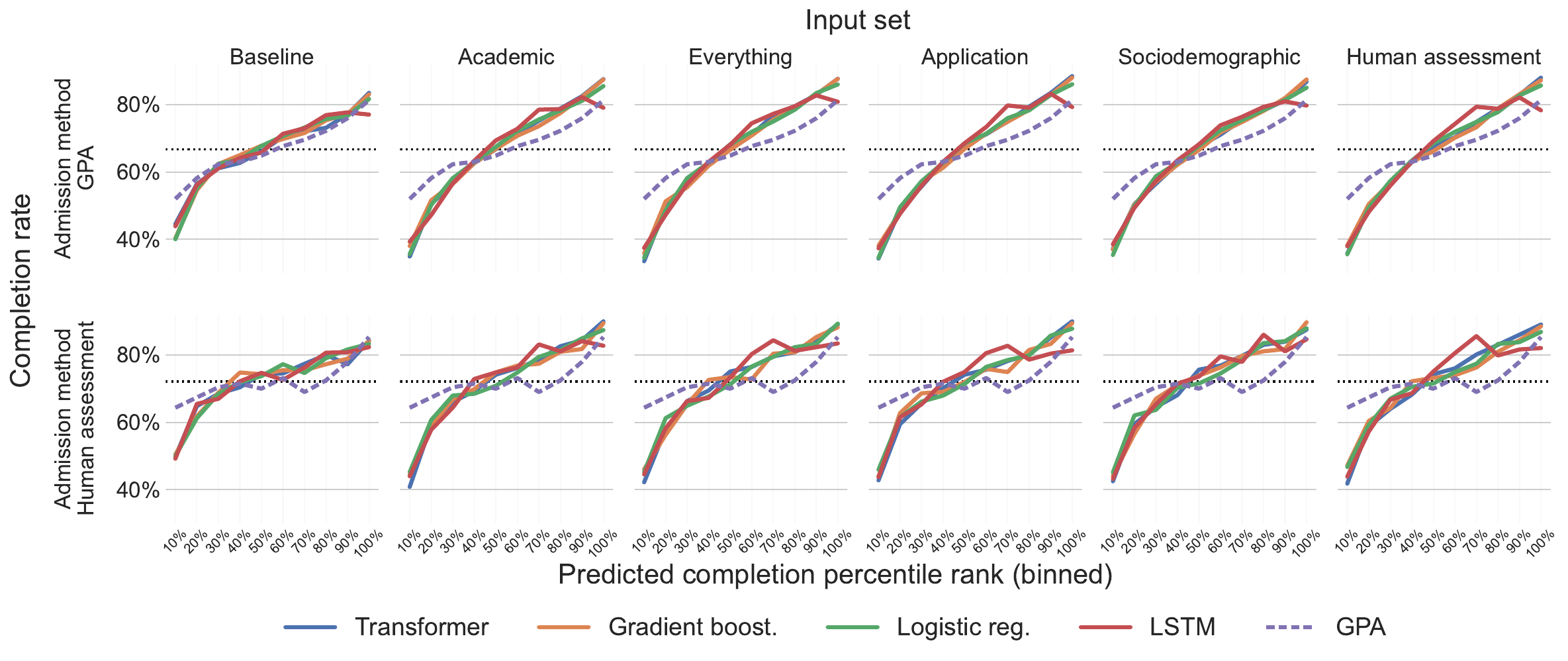}
    \caption[Overall contraction curves across input sets and student populations]{\textbf{Overall contraction curves across models and student populations.} Students' program completion probabilities are binned in ascending order, and mean completion rates are computed for each bin. The top row presents results for students admitted through GPA-based admissions, while the bottom row displays results for those admitted via human assessment. Each column corresponds to a different input set. Perfect ranking would occur if completion rates started at 0\% and then abruptly jumped to 100\%. Dotted lines represent the mean completion rate for each sample.} 
    \label{fig:contract_appendix_ungrouped}
\end{figure}

\begin{figure}[H]
    \centering
        \includegraphics[width=\linewidth]{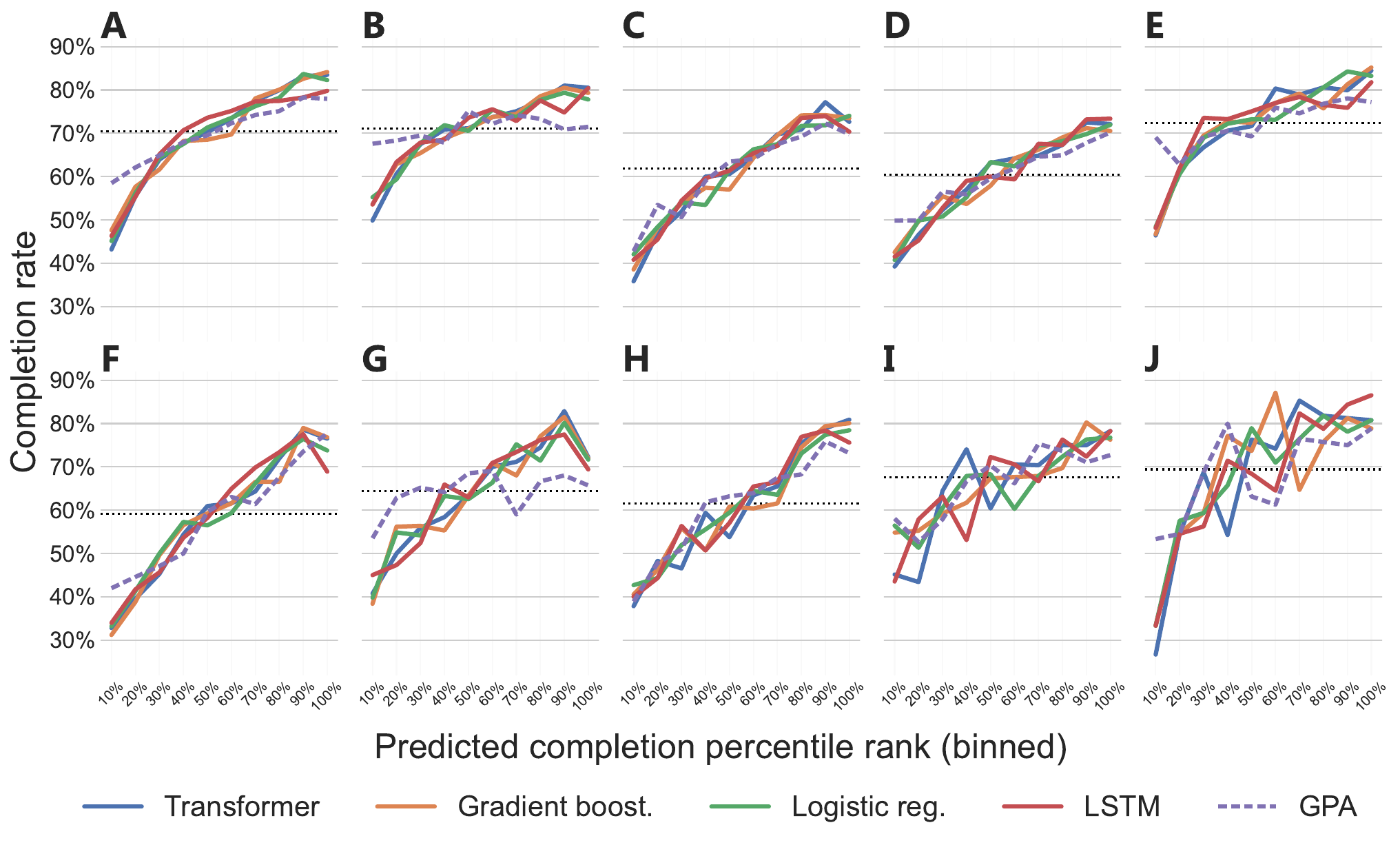}
    \caption[Field-specific study program contraction curves]{\textbf{Field-specific study program contraction curves.} Students' completion probabilities within each study program are binned in ascending order, with mean completion rates computed for each bin. This analysis is conducted for students admitted through GPA-based admissions using the \emph{academic} input set. Panels correspond to different fields, sorted from largest to smallest intake: 
    (A)~Business, administration and law, 
    (B) Health and welfare, 
    (C)~Engineering, manufacturing and construction, 
    (D)~Arts and humanities, 
    (E)~Social sciences, journalism and information, 
    (F)~Natural sciences, mathematics and statistics, 
    (G)~Education
    (H)~Information and communication technologies (ICTs), 
    (I)~Services, 
    (J)~Agriculture, forestry, fisheries, and veterinary.
    } 
    \label{fig:contract_appendix_field_specific}
\end{figure}

\begin{figure}
    \centering
        \centering
        \includegraphics[width=1\linewidth]{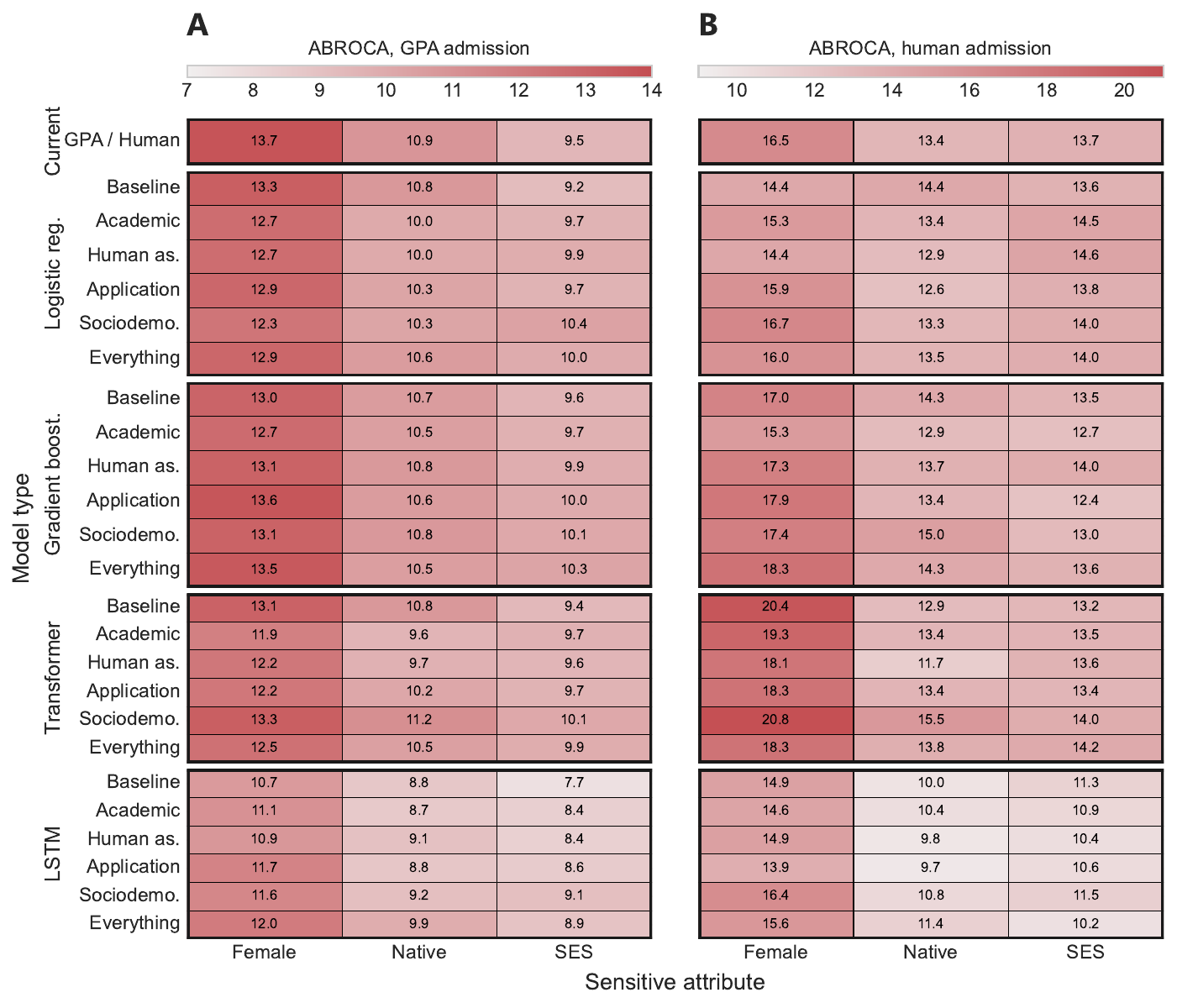}
    \caption[ABROCA scores across models and current admission systems]{\textbf{ABROCA scores across models and current admission systems.} Fairness measures are calculated using the integral described in Section~\ref{sec:fairness_methods}. An average Absolute Between-ROC Area (ABROCA) score weighted by student intake across all institutions is reported. Panel~A presents the results for students admitted through GPA-based admission, while Panel~B shows the results for those admitted through human assessment. The sensitive attributes are whether a student is native (\textit{Native}, not a first- or second-generation immigrant), sex (\textit{Female}), and whether the student is above or below median socioeconomic status (\textit{SES}).}
    \label{fig:abroca_appendix}
    
\end{figure}

\begin{figure}
    \centering
        \centering
        \includegraphics[width=1\linewidth]{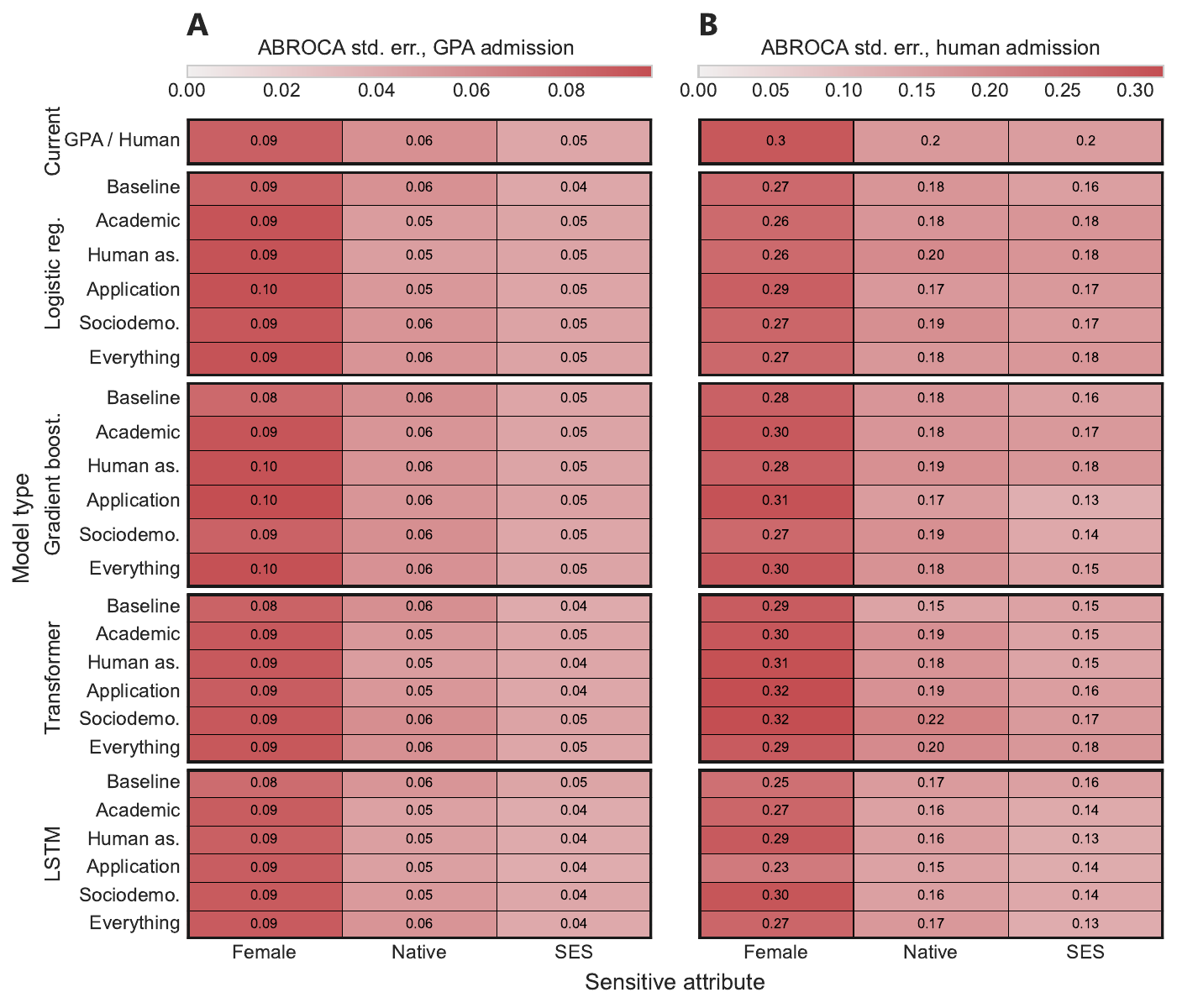}
    \caption[Weighted standard error of ABROCA scores across models and current admission systems]{\textbf{Weighted standard error of ABROCA scores across models and current admission systems.} Fairness measures are calculated using the integral described in Section~\ref{sec:fairness_methods}. A weighted standard error of the Absolute Between-ROC Area (ABROCA) score, weighted by student intake across all institutions, is reported. Panel~A presents the results for students admitted through GPA-based admission, while Panel~B shows the results for those admitted through human assessment. The sensitive attributes are whether a student is first- or second-generation immigrant (\textit{Native}), sex (\textit{Female}), and whether the student is above or below median socioeconomic status (\textit{High SES}).}
    \label{fig:abroca_appendix_standard_errors}
    
\end{figure}

\begin{figure}
    \centering
    \includegraphics[width=1\columnwidth]{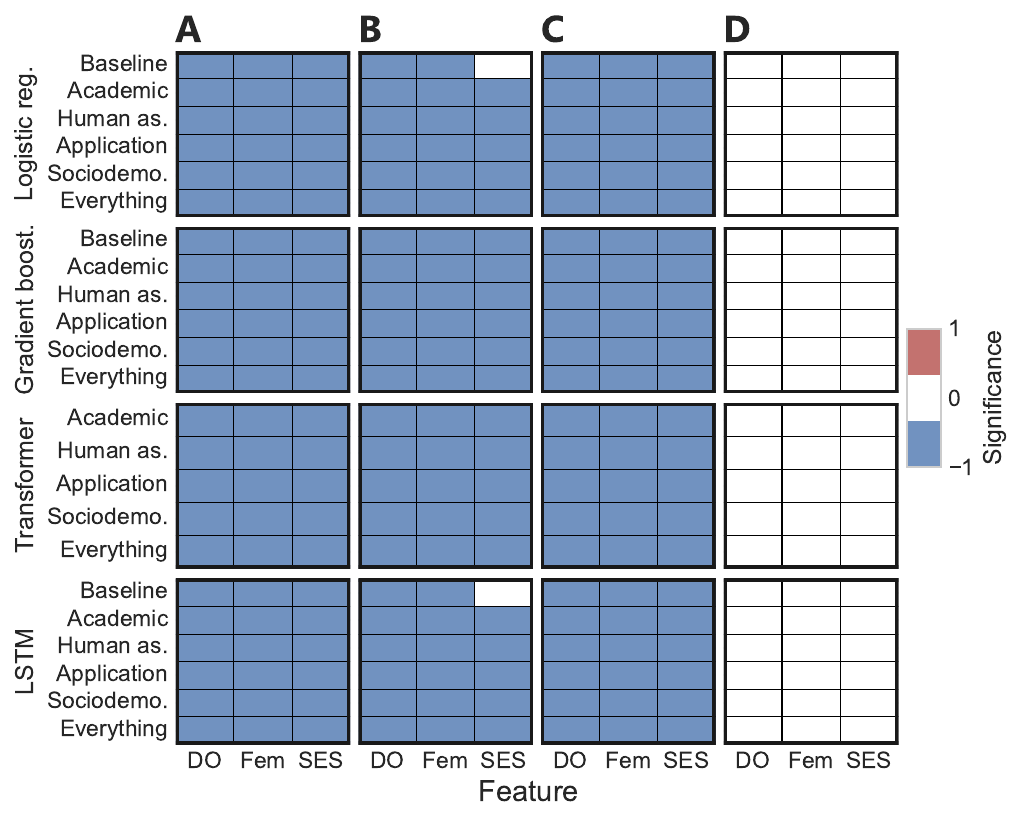}
    \vspace{-0.5cm}
    \caption[Fairness results from $z$-tests for all models]{\textbf{Fairness results from $z$-tests for all models.} The fairness tests reported are the $z$-tests described in Section~\ref{sec:fairness_methods}. Panel~A tests for binary independence, Panel~B for equal false positive rates, Panel~C for equal true positive rates, and Panel~D for sufficiency. A significant positive (negative) test indicates a significantly lower mean in the reference (non-reference) group. The sensitive attributes are whether a student is a native of Danish origin (\emph{DO}: \emph{Danish origin}), sex (\emph{Fem}: \emph{Female}), and whether the student is above or below median socioeconomic status (\emph{SES}: \emph{High SES})}
    \label{fig:fairness_appendix}
\end{figure}

\subsection{Sequence Models }\label{sec:transformer-variables}

\begin{table}[H]
    \centering
    \resizebox{\textwidth}{!}{\begin{tabular}{lccccc}
\toprule
Model & LSTM \# Param. & Transformer \# Param. & Sequence length & \# Channels  & Vocab size \\
\midrule
Academic & 0.5M & 26.1M & 78 & 13 & 1773 \\
Human & 0.5M & 26.1M & 78 & 14 & 1783 \\
Application & 0.5M & 26.2M & 80 & 16 & 2003 \\
Sociodemography & 0.8M & 27.5M & 88 & 17 & 4529 \\
Everything & 0.9M & 27.7M & 90 & 21 & 4778 \\
Baseline & 0.3M & 25.5M & 3 & 1 & 528 \\
\bottomrule
\end{tabular}
}  
    \caption[Model information for sequence models]{\textbf{Model information for sequence models.} \# Parameters include all trainable model parameters. \# Channels includes the channel containing only the \texttt{[CLS]} token. \label{tab:transformer_model_info}
    }
\end{table}

\begin{table}[H]
    \centering
    \resizebox{\textwidth}{!}{\begin{tabular}{lccccccc}
\hline
Variable                          & Type & \textit{Baseline} & \textit{Academic}   & \textit{Application}  &  \textit{Human assessment} &  \textit{Socio-demographic}&  \textit{Everything} \\
\hline
\multicolumn{7}{l}{\textbf{Temporal}} \\
\hspace{1em} Relative year  & Context  & \checkmark & \checkmark & \checkmark & \checkmark & \checkmark & \checkmark  \\

\multicolumn{7}{l}{\textbf{Courses}} \\
\hspace{1em} Grade  & Event & & \checkmark & \checkmark & \checkmark & \checkmark & \checkmark  \\
\hspace{1em} Course  & Context & & \checkmark & \checkmark & \checkmark & \checkmark & \checkmark \\
\hspace{1em} Course type  & Context & &  \checkmark & \checkmark & \checkmark & \checkmark & \checkmark \\
\hspace{1em} Course level &  Context & &  \checkmark & \checkmark & \checkmark & \checkmark & \checkmark \\
\hspace{1em} Test type &  Context  & &  \checkmark & \checkmark& \checkmark & \checkmark & \checkmark  \\
\hspace{1em} Type of education  & Context & & \checkmark & \checkmark & \checkmark & \checkmark & \checkmark  \\
\hspace{1em} Study line &  Context  & \checkmark & \checkmark & \checkmark & \checkmark & \checkmark & \checkmark  \\
\hspace{1em} Primary or high school & Context  & & \checkmark & \checkmark  & \checkmark &  \checkmark & \checkmark \\
\hspace{1em} Institutional ID &  Context  & & &  & &  \checkmark & \checkmark \\

\multicolumn{7}{l}{\textbf{Higher education}} \\
\hspace{1em} GPA\textsuperscript{a}  & Context & \checkmark\textsuperscript{d} & \checkmark & \checkmark & \checkmark & \checkmark & \checkmark \\
\hspace{1em} Place of enrollment, ID & Event & \checkmark & \checkmark & \checkmark & \checkmark & \checkmark & \checkmark  \\
\hspace{1em} Place of enrollment, ISCED broad group & Context &  & \checkmark & \checkmark & \checkmark & \checkmark & \checkmark \\
\hspace{1em} Place(s) of application, ID & Event & &  & \checkmark  & &  & \checkmark \\
\hspace{1em} Place(s) of application, ISCED broad group & Context  & &  &  \checkmark& &  & \checkmark  \\
\hspace{1em} Place(s) of application, rank & Context & &  & \checkmark & &  & \checkmark \\
\hspace{1em} Applied through Quota 2 & Context & &  & \checkmark &  &  & \checkmark \\
\hspace{1em} Enrolled\textsuperscript{b}  & Context & &  & \checkmark & & & \checkmark \\
\hspace{1em} Quota 2 ranking, decile   & Context & & &  & \checkmark &  & \checkmark \\
\hspace{1em} GPA cut-off in previous year    & Context &    &  &  & & \checkmark & \checkmark  \\

\multicolumn{7}{l}{\textbf{Demographic}} \\
\hspace{1em} Age    & Context    & &  &  & & \checkmark & \checkmark  \\
\hspace{1em} Female & Event &  & &  & & \checkmark& \checkmark \\
\hspace{1em} Danish origin & Event & &  &  &  & \checkmark & \checkmark  \\

\multicolumn{7}{l}{\textbf{Sociodemography of parents}\textsuperscript{c}} \\
\hspace{1em} Income & Event & &  &  & & \checkmark& \checkmark \\
\hspace{1em} Wealth & Event & &  &  & & \checkmark& \checkmark \\
\hspace{1em} Education, ISCED code & Event  & &  &  & & \checkmark& \checkmark \\
\hspace{1em} Education, months & Event & &  &  & & \checkmark& \checkmark \\
\hline
\end{tabular}}
    \caption[Variables present in sequence model variants]{\textbf{Variables present in sequence model variants.} The variables are grouped according to the type of information and further classified into either an \textit{event} (i.e., creating a new event in sequence) or \textit{context} (i.e., providing further context to an event). The ISCED broad groups of education follow the ISCED-F 2013 classification, and ISCED codes for education follow the ISCED-2011 classification \cite{eurostat_international_2023}. Footnotes: \textit{a})~GPA is calculated based on high school grades but is included in the higher education events as context, not as a course event (hence the placement). \textit{b})~Due to the inclusion of full application information, a variable is introduced to encode the place of enrollment. This variable is unnecessary when only the place of enrollment is included in the sequence. \textit{c})~Parental information is recorded for each parent. \textit{d})~GPA is included as an event in the baseline model. \label{tab:variables_transformer}}
\end{table}

\subsection{Tabular Models}\label{sec:baseline_models}

\begin{table}[H]
    \centering
    \resizebox{\textwidth}{!}{\begin{tabular}{lccccccc}
\hline
Variables                           & Type & \textit{Baseline} & \textit{Academic}   & \textit{Application}  &  \textit{Human assessment} &  \textit{Socio-demographic}&  \textit{Everything} \\
\hline
\multicolumn{6}{l}{\textbf{Courses}} \\
\hspace{1em} GPA  & Continuous & \checkmark & \checkmark & \checkmark & \checkmark & \checkmark & \checkmark \\
\hspace{1em} Course grade avg.  &  Continuous &  & \checkmark & \checkmark & \checkmark & \checkmark & \checkmark \\
\hspace{1em} Course field grade avg.  &  Continuous &  & \checkmark & \checkmark & \checkmark & \checkmark & \checkmark \\
\hspace{1em} Course field grade std.dev.   & Continuous &  & \checkmark & \checkmark & \checkmark & \checkmark & \checkmark \\
\hspace{1em} Course field number of grades   & Continuous &  & \checkmark & \checkmark & \checkmark & \checkmark & \checkmark \\
\hspace{1em} Type of education & Nominal  &  & \checkmark & \checkmark & \checkmark & \checkmark & \checkmark  \\
\hspace{1em} Study line  & Nominal &  & \checkmark & \checkmark & \checkmark & \checkmark & \checkmark  \\
\hspace{1em} Study program ID  & Nominal &  & &  & &  \checkmark & \checkmark \\

\multicolumn{6}{l}{\textbf{Higher education}} \\
\hspace{1em} Place of enrollment, ID & Nominal &   \checkmark  & \checkmark & \checkmark & \checkmark & \checkmark & \checkmark  \\
\hspace{1em} Place of enrollment, ISCED broad group  & Nominal &  & \checkmark & \checkmark & \checkmark & \checkmark & \checkmark \\\hspace{1em} Place of enrollment, rank  & Ordinal &  & &\checkmark &  & \checkmark & \checkmark  \\
\hspace{1em} Applied through Quota 2 & Binary &  &  & \checkmark &  &  & \checkmark \\
\hspace{1em} Quota 2 ranking, decile  & Ordinal &  &  & & \checkmark & & \checkmark \\
\hspace{1em} GPA cut-off in previous year  & Continuous &  &  &  & & \checkmark& \checkmark  \\
\multicolumn{6}{l}{\textbf{Demographic}} \\
\hspace{1em} Age      & Continuous    &  &  &  & & \checkmark & \checkmark  \\
\hspace{1em} Female  & Binary  &  &  &  & & \checkmark& \checkmark \\
\hspace{1em} Danish origin  & Binary  &  &  &  &  & \checkmark & \checkmark  \\
\multicolumn{6}{l}{\textbf{Sociodemography of parents}*} \\
\hspace{1em} Income   & Continuous &  &  &  & & \checkmark& \checkmark \\
\hspace{1em} Wealth   & Continuous &  &  &  & & \checkmark& \checkmark \\
\hspace{1em} Education, ISCED code & Nominal   &  &  &  & & \checkmark& \checkmark \\
\hspace{1em} Lengths of studies in months & Continuous   &  &  &  & & \checkmark & \checkmark \\
\hline
\end{tabular}}  
    \caption[Variables present in tabular model variants]{\textbf{Variables present in tabular model variants.} Ordinal variables are one-hot encoded in the logistic regression but are included as a single ordinal variable in XGBoost. Parental information is included for each parent. Study program, type of education, and institutional ID of courses are inferred by identifying the most commonly associated type and program for all secondary school grades received. The ISCED broad groups of education follow the ISCED-F 2013 classification and ISCED codes for education follow the ISCED-2011 classification \cite{eurostat_international_2023}. \label{tab:variables_baseline}
    }
\end{table}

\begin{table}[ht]
\centering
\begin{tabular}{lc}
\toprule
Parameter & Distribution \\
\midrule
\textbf{XGBoost}& \\
\hspace{1em} \texttt{learning\_rate} & \texttt{Uniform(0.01, 0.99)} \\
\hspace{1em} \texttt{max\_depth} & \texttt{RandInt(2, 13)} \\
\hspace{1em} \texttt{subsample} & \texttt{Uniform(0.01, 0.99)} \\
\hspace{1em} \texttt{colsample\_bytree} & \texttt{Uniform(0.01, 0.99)} \\
\hspace{1em} \texttt{lambda} & \texttt{Uniform($1e^{-9}$, 100)} \\
\hspace{1em} \texttt{alpha} & \texttt{Uniform($1e^{-9}$, 100)} \\
\hspace{1em} \texttt{n\_estimators} & \texttt{RandInt(50, 5001)} \\
\textbf{Logistic regression} & \\
\hspace{1em} \texttt{C} & \texttt{LogUniform($1e^{-6}$, $1e^{6}$)} \\
\hspace{1em} \texttt{penalty} & \texttt{\{None, l2, l1, elasticnet\}} \\
\hspace{1em} \texttt{l1\_ratio} & \texttt{Uniform(0.01, 0.99)} \\
\bottomrule
\end{tabular}
\caption[Hyperparameter search space for tabular models]{\textbf{Hyperparameter search space for tabular models.} Distributions are specified as in \texttt{scipy.stats}.}
\label{tab:baseline_hyperparams}
\end{table}

\FloatBarrier
\newpage

\subsection{Summary Statistics}

\begin{table}[H]
  
    \begin{tabular}{lll}
 & \multicolumn{1}{c}{2006-2016}& \multicolumn{1}{c}{2017} \\
\hline
 Educational field                                 & Count   & Count   \\
\hline
 Health and welfare                                & 92,379  & 12,001  \\
 Business, administration and law                  & 76,757  & 9,961   \\
 Arts and humanities                               & 57,657  & 5,872   \\
 Social sciences, journalism and information       & 41,027  & 5,611   \\
 Engineering, manufacturing and construction       & 37,821  & 4,577   \\
 Natural sciences, mathematics and statistics      & 24,478  & 2,923   \\
 Education                                         & 22,860  & 2,662   \\
 Information and communication technologies (ICTs) & 13,987  & 2,265   \\
 Services                                          & 6,147   & 1,036   \\
 Agriculture, forestry, fisheries  and veterinary  & 3,538   & 480     \\
 Generic programmes and qualifications             & 151     & 16      \\
\midrule
 Total                                             & 376,802 & 47,404  \\
\hline
\end{tabular}
    \vspace{-0.5cm}
    \caption[Distribution of students across educational fields.]{\textbf{Distribution of students across educational fields}  \label{tab:field_count}}

\end{table}

\begin{table}[H]
    \centering
    \begin{tabular}{llrrlrr}
& \multicolumn{3}{c}{2006-2016}& \multicolumn{3}{c}{2017} \\
\hline
                           & Count   &   Mean &   S.D. & Count   &   Mean &   S.D. \\
\hline
\multicolumn{5}{l}{\textbf{Means}} \\
\hspace{1em} Primary school, languages & 376,782 &   6.54 &   2.33 & 47,397  &   6.78 &   2.44 \\
\hspace{1em} Primary school, other     & 376,793 &   7.26 &   2.14 & 47,403  &   7.48 &   2.31 \\
\hspace{1em} Primary school, STEM      & 376,746 &   6.51 &   2.7  & 47,404  &   6.72 &   2.82 \\
\hspace{1em} High school, languages    & 376,668 &   7.35 &   1.68 & 47,392  &   7.48 &   1.98 \\
\hspace{1em} High school, other        & 362,136 &   8.09 &   2.31 & 46,914  &   7.83 &   2.31 \\
\hspace{1em} High school, STEM         & 376,456 &   7.43 &   1.97 & 47,349  &   7.54 &   2.31 \\
\multicolumn{5}{l}{\textbf{Standard deviation}} \\
\hspace{1em} Primary school, languages & 376,767 &   2.25 &   0.8  & 47,393  &   2.28 &   0.82 \\
\hspace{1em} Primary school, other     & 376,308 &   2.23 &   0.72 & 47,395  &   2.26 &   0.76 \\
\hspace{1em} Primary school, STEM      & 370,302 &   2.19 &   0.93 & 46,335  &   2.2  &   0.98 \\
\hspace{1em} High school, languages    & 376,641 &   2.01 &   0.5  & 47,386  &   2.09 &   0.53 \\
\hspace{1em} High school, other        & 258,695 &   1.79 &   1.09 & 44,846  &   1.88 &   0.97 \\
\hspace{1em} High school, STEM         & 376,392 &   1.77 &   0.59 & 47,345  &   1.83 &   0.58 \\
\multicolumn{5}{l}{\textbf{Counts}} \\
\hspace{1em} Primary school, languages & 376,782 &   7.51 &   2.41 & 47,397  &   7.23 &   2.06 \\
\hspace{1em} Primary school, other     & 376,793 &  10.52 &   3.78 & 47,403  &  10.27 &   3.81 \\
\hspace{1em} Primary school, STEM      & 376,746 &   7.17 &   3.41 & 47,404  &   6.75 &   3.15 \\
\hspace{1em} High school, languages    & 376,668 &  18.56 &   6.2  & 47,392  &  18.77 &   5.79 \\
\hspace{1em} High school, other        & 362,136 &   3.3  &   1.87 & 46,914  &   4.61 &   1.25 \\
\hspace{1em} High school, STEM         & 376,456 &  11.35 &   3.1  & 47,349  &  11.87 &   2.77 \\
\hline
\end{tabular}
    \vspace{-0.5cm}
    \caption[Distribution of course aggregate measures]{\textbf{Distribution of course aggregate measures.} Means, standard deviations, and counts are calculated for each student separately.}
    \label{tab:course_aggregates}
\end{table}

\begin{table}[H]
    \centering
    \begin{tabular}{llr}
\hline
                               & 2006-2016   &   2017 \\
\hline
\multicolumn{3}{l}{\textbf{Tertiary school information}} \\
\hspace{1em} Study program                 & 1,173       &    842 \\
\hspace{1em} Education field               & 11          &     11 \\
\multicolumn{3}{l}{\textbf{Primary and secondary school information}} \\
\hspace{1em} Types                         & 9           &      9 \\
\hspace{1em} High school specialization    & 2,234       &    815 \\
\hspace{1em} Institutions                  & 2,484       &    316 \\
\multicolumn{3}{l}{\textbf{Course information}} \\
\hspace{1em} Course                        & 139         &     96 \\
\hspace{1em} Course with level             & 272         &    145 \\
\hspace{1em} Course with level \& test type & 649         &    312 \\
\hline
\end{tabular}
    \caption[Number of unique values per categorical feature.]{\textbf{Number of unique values per categorical feature}}
    \vspace{-0.5cm}
    \label{tab:categorical}
\end{table}

\begin{table}[H]
    \centering
    \resizebox{\textwidth}{!}{\begin{tabular}{lllllll}
& \multicolumn{3}{c}{2006-2016}& \multicolumn{3}{c}{2017} \\
\hline
                          & Count   & Mean    & S.D.      & Count   & Mean    & S.D.      \\
\hline
\multicolumn{5}{l}{\textbf{Educational}} \\
\hspace{1em} Graduated                & 376,802 & 0.69    & 0.46      & 47,404  & 0.68    & 0.47      \\
\hspace{1em} GPA                      & 376,136 & 7.08    & 2.28      & 47,345  & 7.22    & 2.33      \\
\hspace{1em} Acceptance priority      & 376,802 & 1.47    & 4.26      & 47,404  & 1.30    & 0.82      \\
\hspace{1em} Applied through quota 2  & 376,802 & 0.32    & 0.47      & 47,404  & 0.38    & 0.49      \\
\hspace{1em} Accepted through quota 2 & 376,802 & 0.14    & 0.35      & 47,404  & 0.19    & 0.39      \\
\hspace{1em} GPA cutoff previous year & 342,125 & 4.22    & 2.06      & 44,822  & 4.66    & 2.20      \\
\multicolumn{5}{l}{\textbf{Sociodemographic}} \\
\hspace{1em} Age                      & 371,717 & 20.35   & 1.88      & 46,760  & 20.99   & 2.40      \\
\hspace{1em} Female                   & 376,802 & 0.58    & 0.49      & 47,404  & 0.56    & 0.50      \\
\hspace{1em} Danish origin            & 376,802 & 0.91    & 0.29      & 47,404  & 0.89    & 0.32      \\
\hspace{1em} Income, mom              & 371,152 & 257,431 & 268,714   & 46,562  & 287,183 & 178,392   \\
\hspace{1em} Income, dad              & 358,664 & 340,308 & 798,540   & 44,892  & 389,321 & 724,069   \\
\hspace{1em} Wealth, mom              & 371,152 & 412,558 & 3,787,772 & 46,562  & 360,095 & 2,462,602 \\
\hspace{1em} Wealth, dad              & 358,664 & 779,645 & 4,054,591 & 44,892  & 658,163 & 3,491,393 \\
\hspace{1em} Education length, mom    & 372,049 & 174.28  & 31.16     & 46,697  & 174.27  & 31.92     \\
\hspace{1em} Education length, dad    & 365,541 & 174.57  & 31.68     & 45,961  & 173.89  & 31.16     \\
\hline
\end{tabular}
}

    \vspace{-0.5cm}
    \caption[Summary statistics of sociodemographic and educational information across time periods.]{\textbf{Summary statistics of sociodemographic and educational information across time periods} }
    \label{tab:summary_stats}
\end{table}

\begin{table}[H]
    \centering
    \resizebox{\textwidth}{!}{\begin{tabular}{lllllll}
& \multicolumn{3}{c}{GPA admission}& \multicolumn{3}{c}{Human admission} \\
\hline
                          & Count   & Mean    & S.D.      & Count   & Mean    & S.D.      \\
\hline
\multicolumn{5}{l}{\textbf{Educational}} \\
\hspace{1em} Graduated                & 38,401  & 0.67    & 0.47      & 9,003   & 0.74    & 0.44      \\
\hspace{1em} GPA                      & 38,383  & 7.57    & 2.27      & 8,962   & 5.76    & 2.02      \\
\hspace{1em} Acceptance priority      & 38,401  & 1.28    & 0.80      & 9,003   & 1.35    & 0.86      \\
\hspace{1em} Applied through quota 2  & 38,401  & 0.24    & 0.43      & 9,003   & 1.00    & 0.00      \\
\hspace{1em} Accepted through quota 2 & 38,401  & 0.00    & 0.00      & 9,003   & 1.00    & 0.00      \\
\hspace{1em} GPA cutoff previous year & 37,458  & 4.54    & 2.20      & 7,364   & 5.26    & 2.11      \\
\multicolumn{5}{l}{\textbf{Sociodemographic}} \\
\hspace{1em} Age                      & 37,885  & 20.81   & 2.36      & 8,875   & 21.79   & 2.43      \\
\hspace{1em} Female                   & 38,401  & 0.54    & 0.50      & 9,003   & 0.62    & 0.48      \\
\hspace{1em} Danish origin            & 38,401  & 0.88    & 0.32      & 9,003   & 0.90    & 0.30      \\
\hspace{1em} Income, mom              & 37,716  & 287,200 & 177,392   & 8,846   & 287,112 & 182,606   \\
\hspace{1em} Income, dad              & 36,385  & 389,322 & 719,954   & 8,507   & 389,314 & 741,457   \\
\hspace{1em} Wealth, mom              & 37,716  & 367,345 & 2,644,687 & 8,846   & 329,183 & 1,448,646 \\
\hspace{1em} Wealth, dad              & 36,385  & 665,821 & 3,295,065 & 8,507   & 625,410 & 4,229,549 \\
\hspace{1em} Education length, mom    & 37,820  & 174.53  & 32.12     & 8,877   & 173.16  & 31.03     \\
\hspace{1em} Education length, dad    & 37,214  & 174.25  & 31.45     & 8,747   & 172.37  & 29.85     \\
\hline
\end{tabular}
}
    
    \vspace{-0.5cm}
    \caption[Summary statistics of sociodemographic and educational information across admission type.]{\textbf{Summary statistics of sociodemographic and educational information across admission type}. The summary statistics are calculated for the sample in the test year of 2017.}
    \label{tab:summary_stats_quota_1_and_2}
\end{table}

\end{appendices}

\end{document}